\documentclass[12pt,a4paper]{article}
\pdfoutput=1
\usepackage{jheppub}
\usepackage{tikz}
\usepackage{subcaption}
\usepackage{natbib}
\usepackage{physics}
\usepackage{braket}

\begin{document}

\begin{flushright}
	\hfill{OU-HET-1218}
\end{flushright}
\title{\mathversion{bold} Krylov complexity as an order parameter for deconfinement phase transitions at large $N$}

 \author[a]{Takanori Anegawa,}
 \author[a]{Norihiro Iizuka}
 \author[b]{and Mitsuhiro Nishida}
\affiliation[a]{\it Department of Physics, Osaka University, Toyonaka, Osaka 560-0043, JAPAN}
\affiliation[b]{\it Department of Physics, Pohang University of Science and Technology, Pohang 37673, Korea}

\emailAdd{takanegawa@gmail.com, iizuka@phys.sci.osaka-u.ac.jp, nishida124@postech.ac.kr}

\abstract{Krylov complexity has been proposed as a diagnostic of chaos in non-integrable lattice and quantum mechanical systems, and if the system is chaotic, Krylov complexity grows exponentially with time. However, when Krylov complexity is applied to quantum field theories, even in free theory, it grows exponentially with time. This exponential growth in free theory is simply due to continuous momentum in non-compact space and has nothing to do with the mass spectrum of theories. Thus by compactifying space sufficiently, exponential growth of Krylov complexity due to continuous momentum can be avoided. In this paper, we propose that the Krylov complexity of operators such as $\mathcal{O}=\Tr[F_{\mu\nu}F^{\mu\nu}]$ can be an order parameter of confinement/deconfinement transitions in large $N$ quantum field theories on such a compactified space. We explicitly give a prescription of the compactification at finite temperature to distinguish the continuity of spectrum due to momentum and mass spectrum. We then calculate the Krylov complexity of $\mathcal{N}=4, 0$ $SU(N)$ Yang-Mills theories in the large $N$ limit by using holographic analysis of the spectrum and show that the behavior of Krylov complexity reflects the confinement/deconfinement phase transitions through the continuity of mass spectrum.}

\maketitle

\section{Introduction}

How to determine whether a quantum system is chaotic is a question that has long been discussed in physics. One traditional characterization of quantum chaos is given by the level spacing statistics of the energy spectrum calculated from the Hamiltonian of quantum systems. In chaotic systems, the energy level spacing is expected to obey a Wigner-Dyson distribution, which is a characteristic behavior in random matrix theories \cite{wigner1951class, Dyson:1962es, berry1977level, Bohigas:1983er}.

Intuitively, chaos means that a small initial change can make a large difference later. Based on how an operator changes under time evolution, one can discuss the quantum chaos of operators, which depends on the choice of operator. There is also ambiguity in the choice of physical quantities to measure how much the operator changes.

An important quantity for quantum chaos dependent on operators is the out-of-time-ordered correlator (OTOC) \cite{larkin1969quasiclassical}. An exponential behavior of the OTOC, quantified by the nonzero Lyapunov exponent, has been proposed as a measure of quantum chaos \cite{Kitaev-talks:2015}. As long as correlation functions can be defined and calculated, the OTOC can be used as the measure even in quantum field theories whose energy spectrum is continuous.

Another important quantity that has been well studied recently for quantum chaos is the Krylov complexity \cite{Parker:2018yvk}, which is a measure of how fast an operator $\mathcal{O}$ spreads in a subspace, called Krylov subspace, of the Hilbert space. This is a quantity that indicates scrambling in the Krylov subspace and generally exhibits either merely oscillatory, linearly increasing, or exponentially increasing behavior. It is conjectured that, in the thermodynamic limit, the Krylov complexity grows exponentially in non-integrable systems.

Since the definitions of Krylov complexity and OTOC are different, the exponential growth of these two measures evaluates different aspects of quantum chaos. The Krylov complexity is a measure of the operator growth\footnote{One can also define the Krylov complexity for the time evolution of states, which is called spread complexity \cite{Balasubramanian:2022tpr}. The spread complexity of chaotic systems whose energy spectrum are described by random matrix theories has been well studied recently \cite{Balasubramanian:2022dnj, Erdmenger:2023wjg,  Bhattacharyya:2020qtd}.} of an operator $\mathcal{O}$ in the Krylov subspace under the time evolution, while the OTOC is a measure of how an operator $\mathcal{O}_1$ affects another operator $\mathcal{O}_2$ via a commutator $[\mathcal{O}_1(t),\mathcal{O}_2(0)]$. Nevertheless, it is conjectured that the following bound $\lambda \leq \alpha$ exists, where $\alpha$ is the exponent in the exponential growth of the Krylov complexity, and $\lambda$ is the Lyapunov coefficient of the OTOC.

There is one fault in the exponential growth of Krylov complexity as a measure of quantum chaos. In a simple non-interacting free scalar quantum field theory on non-compact space, it is shown that the Krylov complexity grows exponentially \cite{Dymarsky:2021bjq}. This exponential growth in a free field theory is due to continuous momentum in non-compact spatial directions. Even though there is a mass gap in the IR region, the spectrum of a scalar operator is continuous due to continuous momentum, and thus the Krylov complexity grows exponentially as shown in free massive scalar theories on non-compact space \cite{Avdoshkin:2022xuw, Camargo:2022rnt}. The continuity of momentum in quantum field theories arises by taking a continuous limit of zero lattice spacing in lattice systems.

A simple solution to this fault is to compactify the space on which a quantum field theory lives. Such an analysis of the Krylov complexity in quantum field theories on compact space was explicitly demonstrated by \cite{Avdoshkin:2022xuw, Kundu:2023hbk}. In particular, from the viewpoint of AdS/CFT \cite{Maldacena:1997re}, the behavior of Krylov complexity under a thermal phase transition dual to the Hawking-Page transition \cite{Hawking:1983tob} was studied.

In holography, the Hawking-Page transition in the bulk can be interpreted as a confinement/deconfinement phase transition in the large $N$ quantum field theory side \cite{Witten:1998zw}. Even though the degrees of freedom of systems are infinite, the spectrum in quantum field theories can be discrete or continuous. A specific example is a discrete spectrum of the confinement phase and a continuous spectrum of the deconfinement phase in a large $N$ QCD-like theory. The computations of Krylov complexity in \cite{Avdoshkin:2022xuw, Kundu:2023hbk} suggest that the behavior of Krylov complexity is sensitive to the confinement and deconfinement phases.

In this paper, we propose that the Krylov complexity can be an order parameter of such a confinement/deconfinement phase transition in large $N$ field theories and specifically study how the Krylov complexity acts as the order parameter. For a concrete proposal, we consider the following free theory that models the spectrum of a two-point function in holographic QCD where particles of various masses exist,
\begin{align}
\label{Intromodel}
S = \int_{\mathbb{S}^1 \times \mathbb{S}^1} d^2x \sum_{n=0}^{\infty}\left( \frac{1}{2}\partial^{\mu} \phi_n \partial_{\mu} \phi_n + \frac{1}{2}m_n^2 \phi_n^2 \right),\ \ m_n = m +n \delta m.
\end{align}
The model consists of an infinite number of scalar fields $\{ \phi_n\}_{n=0}^{\infty}$, where the smallest mass is $m$, and $\delta m$ represents the gap between masses. This theory is on the thermal circle $\mathbb{S}^1$ with inverse temperature $\beta$ and the spatial circle $\mathbb{S}^1$ due to the compactification of space $x = x+L$. We compactify the space to avoid the continuity of momentum in spatial direction.
By compactifying the space sufficiently to ignore nonzero discrete momentum, the discreteness of the spectrum is determined by $\beta\delta m$ only. When $\beta\delta m \gtrsim  1$, the spectrum is discrete due to the gap $\delta m$, and the  Krylov complexity oscillates and does not grow.  When $\beta\delta m \ll 1$, the spectrum is continuous, and the Krylov complexity grows exponentially. This model is treated as example 3 in Section \ref{sec2}.

From the above example, we propose the following prescription. Since we are interested in the confinement/deconfinement phase transition, let us take the temperatures around which the phase transition occurs such as $\beta \sim \Lambda_{\rm QCD}^{-1}$. Next, compactify the space sufficiently as $\beta / L \gtrsim1$ to avoid continuous momentum. Then, the discreteness of the spectrum depends only on the mass spectrum, not on the momentum. In the confinement phase, $\beta\delta m \gtrsim 1$ yields the oscillational behavior of Krylov complexity. In the deconfinement phase, $\beta\delta m \ll1$ yields the exponential growth of Krylov complexity. This allows the Krylov complexity to act as an order parameter of the confinement/deconfinement phase transition in large $N$ field theories.

Finally, we also examine the case of holographic QCD. The essence is the same as in the above model. Specifically, we calculate the spectrum via bulk geometries with and without AdS black holes, where black holes exhibit extremely strong chaos \cite{Sekino:2008he, Shenker:2013pqa,Shenker:2013yza,Shenker:2014cwa, Maldacena:2015waa}. Then, we evaluate the Krylov complexity from the obtained spectrum.

The basic structure of this paper is as follows: in Section \ref{sec2}, we present several examples that support our proposal. Specifically, we review a free scalar field on a sphere with radius $R$ and the IP matrix model as examples where the discreteness of the spectrum changes. The model of \eqref{Intromodel} is particularly important, which reflects the structure of the holographic QCD spectrum that includes particles of various masses. We then propose a prescription for the Krylov complexity to be an order parameter of the confinement/deconfinement phase transition in large $N$ field theories. In Section \ref{sec3}, we study pure $SU(N)$ Yang-Mills theories, which are of most interest to us as systems that specifically cause (de)confinement. We consider the pure $\mathcal{N}=4$ Super Yang-Mills theory and $\mathcal{N}=0$ pure Yang-Mills theory in the large $N$ limit, and how the Krylov complexity behaves by reading their spectrum through the holographic/D-brane picture. Then we show specifically that the behavior of Krylov complexity changes, indicating a phase transition between deconfinement and confinement in the large $N$ quantum field theories.

\section{Several Examples of Krylov complexity and Our Proposal}\label{sec2}
In this section, we specifically review the Krylov complexity with some examples. 
Then, to clarify the discreteness of the spectrum due to mass and momentum, we presents and analyze a model of infinitely many free scalars with various masses in compact space.
Motivated by these examples, we propose that the Krylov complexity can be an order parameter of a confinement/deconfinement phase transition in large $N$ quantum field theories. 

Before the examples, let us define the Krylov complexity \cite{Parker:2018yvk}. Consider a local operator $\mathcal{O}$ and its time evolution $\mathcal{O}(t)=e^{iHt} \mathcal{O}e^{-iHt}$. We expand $\mathcal{O}(t)$ as
\begin{align}\label{OtExpansion}
\mathcal{O}(t)=\sum_{n=0} i^n \varphi_n(t) \mathcal{O}_n, \;\;\; \mathcal{O}_0:=\mathcal{O}\,,
\end{align}
where $\mathcal{O}_n$ is the Krylov basis constructed by the Lanczos algorithm \cite{Lanczos:1950zz}. The Krylov basis $\mathcal{O}_n$ is an orthonormal basis such that
\begin{align}
(\mathcal{O}_m\vert\mathcal{O}_n)=\delta_{mn},
\end{align}
where $(\mathcal{O}_m\vert\mathcal{O}_n)$ is a suitable inner product between $\mathcal{O}_m$ and $\mathcal{O}_n$. The coefficient $\varphi_n(t)$ in the expansion (\ref{OtExpansion}) obeys the following time evolution
\begin{align}
\frac{d\varphi_{n}(t)}{dt}=ia_n \varphi_{n}(t)-b_{n+1}\varphi_{n+1}(t) +b_{n}\varphi_{n-1}(t) \,,\label{recursionwf}
\end{align}
where $a_n$ and $b_n$ are called Lanczos coefficients in the Lanczos algorithm. By using $\varphi_n(t)$, the Krylov complexity $K(t)$ is defined by
\begin{align}\label{DefKC}
K(t):= \sum_{n=1}^{\infty} n |\varphi_n(t)|^2\,.
\end{align}

By using the inner product, let us define a two-point function $G(t)$ as
\begin{align}
G(t):=(\mathcal{O}(t)\vert\mathcal{O})=\varphi_{0}^*(t),
\end{align}
where an example is (\ref{GtHarmonicOscillator}) in Appendix \ref{App0}.
There exists a numerical algorithm to compute the Lanczos coefficients from $G(t)$ \cite{RecursionBook}. Thus, if the two-point function $G(t)$ or its spectrum $G(\omega):= \int dt e^{i\omega t} G(t)$ is given, we can determine the Lanczos coefficients $a_n$ and $b_n$. Then, by solving $\varphi_{n}(t)$ from (\ref{recursionwf}), we can compute the Krylov complexity (\ref{DefKC}). From now, we will explain some examples of $G(t)$ and their Krylov complexity. Please refer to Appendix \ref{AppA} for more details.\\

\noindent {\bf Example 1: A single free scalar}\\
As a first example, let us calculate the Krylov complexity for a massless minimally-coupled scalar theory in 3-dimensional space $\mathbb{S}^3$ with radius $R$ (and consider a more thermal theory, $\mathbb{S}^3\times \mathbb{S}^1$ as a (3+1)-dimensional theory). In Euclidean signature, the action is
\begin{align}
S = \int_{\mathbb{S}^1 \times \mathbb{S}^3} d^4x \sqrt{g} \left( \frac{1}{2}\partial^{\mu} \phi \partial_{\mu} \phi + \frac{\xi}{2}\mathcal{R} \phi^2 \right)
\label{ex1model}
\end{align}
where $\xi$ is the minimal coupling $\xi = \frac{d-2}{4(d-1)}$ in general $d$ dimension and $\mathcal{R}$ is scalar curvature of background $\mathbb{S}^{3}$, and in $d=4$, $\xi = {1}/{6}$, $\mathcal{R} = 6/R^2$. 

If $R$ is not much larger than the inverse temperature $\beta$, then due to the Kaluza-Kelin (KK) tower associated with the compactification in the spatial direction, this corresponds to a typical extension of a mere harmonic oscillator system. However, if $R$ is much larger than $\beta$, then this corresponds to a typical field theory on non-compact space where the behavior of the Krylov complexity changes significantly. This analysis was done in \cite{Dymarsky:2021bjq,Avdoshkin:2022xuw} and we review it here. 

First, we want to obtain a correlation function separated in Euclidean time $\tau$, 
\begin{align}
C(\tau,R) = \langle \phi(\tau,x) \phi(0,x) \rangle_{\beta} \,,
\end{align}
and then the Lorentz version is computed by analytical continuation. By part integration, the Euclidean action is
\begin{align}
S = \int_{\mathbb{S}^1 \times \mathbb{S}^3} d^4x \sqrt{g}\frac{1}{2} \phi \left( \hat{D} + \xi \mathcal{R} \right)\phi \,,\qquad \hat{D} \equiv -\partial_{\tau}^2 -\nabla_{\mathbb{S}^3}^2 \,.
\end{align}
By using the heat kernel method, we can express
\begin{align}
C(\tau,R)&= \braket{\tau,x| \frac{1}{\hat{D} + \xi \mathcal{R}}|0,x} 
= \int_0^{\infty} ds \braket{\tau,x|e^{-s(\hat{D} +\xi \mathcal{R})}|0,x} \nonumber \\
&= \int_0^{\infty} ds K(s,\tau)e^{-s/R^2},
\end{align}
where we use 
$\xi = \frac{1}{6}, \mathcal{R} = \frac{6}{R^2}$  and $K(s,\tau)$ is heat kernel in $\mathbb{S}^1 \times \mathbb{S}^3$, 
\begin{align}
K(s,\tau) &= \braket{\tau,x|e^{-s\hat{D}}|0,x}.
\end{align}
Then we can decompose
\begin{align}
K(s,\tau) &= K_{\mathbb{S}^1}(s,\tau) \times K_{\mathbb{S}^3}(s,x,x),
\end{align}
where 
\begin{align}
K_{\mathbb{S}^1}(s,\tau) & =  \frac{1}{\sqrt{4\pi s}} \sum_{n=-\infty} ^{\infty} e^{-\frac{(\tau+n\beta)^2}{4s}},\\
K_{\mathbb{S}^3}(s,x,x) &= \frac{e^{s/R^2}}{(4\pi s)^{3/2}} \sum_{\ell=-\infty} ^{\infty} e^{-\frac{\pi^2 R^2 \ell^2}{s}}\left( 1- 2 \frac{\pi^2 R^2 \ell^2}{s}\right).
\end{align}
See Appendix \ref{App1} for more details.
From these,
\begin{align}
C(\tau,R)&= \sum_{n,\ell=-\infty} ^{\infty} \int ds \frac{1}{(4\pi s)^2} e^{-\frac{(\tau+n\beta)^2+(2\pi R \ell)^2}{4s}}\left( 1- 2 \frac{\pi^2 R^2 \ell^2}{s}\right),
\end{align}
where $n$ corresponds to the KK tower associated with the compactification in the Euclid $\mathbb{S}^1$ direction and $\ell$ corresponds to that in the $\mathbb{S}^3$ direction.\\
Integrating this and rescaling $\tau \to \beta \tau$ and $R \to \frac{\beta}{2\pi} R$ and performing translation $\tau \to \tau+1/2$ to make this Wightman inner product two-point function. Then we can obtain 
\begin{align}
C(\tau+1/2,R)&\propto \sum_{n,\ell \in \mathbb{Z}} \frac{(\tau+1/2+n)^2-(R\ell)^2}{((\tau+1/2+n)^2+(R\ell)^2)^2}\ \ \quad (0\leq \tau\leq 1) \\
&= \frac{\pi^2}{R^2}\sum_{n\in \mathbb{Z}} \frac{1}{\sinh^2 ((n+1/2+\tau)\pi/R)} \,. 
\end{align}

If $R$ is much smaller than $\beta=1$, then the correlator can be approximated as follows
\begin{align}
C(\tau+1/2,R) \sim \frac{\pi^2}{R^2}\left( \frac{1}{\sinh^2 ((1/2+\tau)\pi/R)} + \frac{1}{\sinh^2 ((-1/2+\tau)\pi/R)} \right).
\end{align}
By using the Toda chain method, we can compute Lanczos coefficients 
where the Lanczos coefficients are divided into odd and even branches. 
\begin{align}\label{twoslope}
b_n^2 = \left( \frac{2\pi}{R}\right)^2
\begin{cases}
(n+1)^2/4 & n=1,3,\cdots \\
\frac{4n(n+1)^2}{n+2}e^{-\pi/R} & n=2,4,\cdots \\
\end{cases}
\end{align}
This behavior is the behavior reproduced when the spectrum is a set of delta functions, and the Krylov complexity exhibits oscillatory behavior.

In an opposite case,  if $R$ is much larger than $\beta=1$, only $\ell=0$ is dominant, and asymptotic behaviors of the correlator and spectrum density are
\begin{align}
C(\tau+1/2,R)&\sim \sum_{n \in \mathbb{Z}} \frac{1}{(\tau+1/2+n)^2}=\frac{\pi^2}{\cos^2(\pi \tau)} =  \frac{\pi^2}{\cosh^2\left(\frac{\pi}{\beta} t \right)} \,,\\
f(\omega) &\sim \sqrt{\frac{\pi}{2}}\beta^2 \omega \csch \left(\frac{\beta \omega}{2}\right) \sim e^{- \frac{\beta \omega}{2}}.
\end{align}
The relationship between the asymptotic form of the spectrum density and the Lanczos coefficient can be identified as follows. 
\begin{align}
\lim_{n \to \infty} b_n=\frac{\pi }{\beta}\, n
\end{align}
Now the asymptotic behavior of $b_n$ is linear, we can find this Krylov complexity shows exponential growth. The asymptotic behavior is
\begin{align}
K(t) \propto e^{ 2\pi t/\beta}\ .
\end{align}
In particular, the exponent is $\alpha=\frac{2\pi }{\beta}$ and is certainly bound to the actual quantum Lyapunov exponent $\lambda=0$ of the free theory. Note, however, that this is a ``bad'' bound since we want the exponent of the Krylov complexity to be zero if the theory is free and thus non-chaotic.

Of particular note is that this is very closely related to the compactness of space. If the space is compact, the momentum in spatial directions is generally discretely quantized. 
Then through $E^2 = m^2 + \vec{k}^2$, 
the energy spectrum $E$ is no longer continuous for the case of gapped mass spectrum $m$.
On the other hand, if the space is noncompact, momentum $\vec{k}$ is always continuous, and thus energy spectrum $E$ is also continuous even in the case of gapped mass spectrum $ m$. In this way, there is a noticeable difference in the energy spectrum dependent on the compactness of the space. 
Since we consider a finite temperature system, the compactness of space is measured by a ratio between $\beta$ and $R$.

The conclusion is that if the radius of the sphere obeys $R \lesssim \beta$, $K(t)$ shows just oscillation behavior. However, if we consider the case of $R \gg \beta$, i.e., the limit to flat space, we reproduce the exponential growth of $K(t)$ even though the theory \eqref{ex1model} is free. Especially in Fig.~3 of \cite{Avdoshkin:2022xuw}, where specific numerical calculations are made, and it is found that the Krylov complexity indeed exhibits oscillatory behavior when $R\sim \beta$. On the contrary, when $R/\beta\to\infty$, the Krylov complexity shows exponential growth.\\

Another important work was done in \cite{Kundu:2023hbk} as follows. Let us consider a two-dimensional holographic CFT, which is dual to AdS$_3$ gravity, on a cylinder $\mathbb{R}^1 \times \mathbb{S}^1$, where $\mathbb{S}^1$ is a compact space. They showed that by varying the scaling dimension of a primary state to define an inner product, the Krylov complexity exhibits a transition of its behavior between oscillation and exponential growth. Since the heavy primary state corresponds to a black hole geometry in AdS/CFT, this transition of the Krylov complexity means that the Krylov complexity can capture the Hawking-Page transition in the bulk. This is also a reflection of the change in the spectrum from discrete to continuous in terms of the bulk fields, due to the going-boundary conditions on the black hole geometry. They also showed that the Krylov complexity in two-dimensional free and Ising CFTs does not exhibit such a transition.\\

\noindent{\bf Example 2: IP model}\\
Next, we consider a quantum mechanical large $N$ matrix model called the IP model \cite{Iizuka:2008hg}. Specifically, the Hamiltonian of the IP model is given by
\begin{align}
H=\frac{1}{2}\Tr (\Pi^2)+\frac{m^2}{2}\Tr (X^2)+\pi^{\dagger}(1+gX/M)\pi+M^2 \phi^{\dagger}(1+g X/M)\phi.
\end{align}
Here, $X_{ij}$ ($U(N)$ adjoint representation) and $\phi_i$ ($U(N)$ fundamental representation) are harmonic oscillator variables, and $\Pi_{ij}$ and $\pi_i$ are those conjugate momentum. 
We consider the following two-point function $G(t)$
\begin{align}
e^{iMt}\langle {\rm T} a_i(t) a_j^{\dagger}(0)\rangle_\beta\equiv \delta_{ij}G(t),
\end{align}
where $a_i^{\dagger}$ and $a_i$ are creation/annihilation operator for the fundamental field $\phi_i$. In the large $N$ and $M$ limit, one can solve the spectrum of $G(t)$ by using the Schwinger-Dyson equation. Then, the Lanczos coefficients and the Krylov complexity can be evaluated from the spectral density $F(\omega):=\Re G(\omega)/\pi$. Varying temperature $T$ and adjoint mass $m$, the Krylov complexity $K(t)$ exhibits various behaviors.
A brief summary is as follows \cite{Iizuka:2023pov,Iizuka:2023fba}.\\

\noindent{\bf Massless case $m=0$ }\\
In this case, the spectrum density is given by a single Wigner semicircle, which is a bounded continuous spectrum. The Krylov complexity shows a linear increase with respect to time $t$.\\

\noindent{\bf Nonzero Mass case $m\ne0$}\\
At zero temperature $T=0$ with nonzero mass, the spectral density can be solved analytically and given by a collection of the delta function, which is a discrete spectrum. At infinite temperature $T\to\infty$ with nonzero mass, the spectral density is a continuous spectrum whose asymptotic behavior can be solved analytically. 
From these spectra, the Krylov complexity can be calculated. In the zero temperature case, the Krylov complexity just oscillates and does not grow due to the discrete spectrum.
In the high-temperature limit, the Krylov complexity grows exponentially with respect to $\sqrt{t}$.\\

What is important to note is that in the IP model with nonzero adjoint mass, the behavior of spectral density changes from discrete to continuous by raising the temperature from zero to nonzero, which corresponds to a phase transition in the large $N$ limit from a confinement phase to a deconfinement phase. Accordingly, the Krylov complexity also changes from just oscillatory to exponentially increasing.

In the massless adjoint case, {\it i.e.}, when the spectral density is given by a single Wigner semicircle, the Krylov complexity grows linearly rather than exponentially. The spectral density of the massless case is continuous, but there is an upper bound in the spectral density, which causes the qualitative change of Krylov complexity. 
In other words, it can be inferred that for the exponential growth of Krylov complexity, it is not only important that the spectral density is continuous, but also that there is no upper bound in the spectral density. \\

Before going to the next example, we would like to speculate on the general situation when the Krylov complexity increases exponentially.

\noindent From the above examples, the following can be deduced. In general, the discrete  spectrum shows an oscillatory behavior of the Krylov complexity. For example, the IP model with non-zero adjoint mass at zero temperature, a single free scalar in compact space 
that exhibit the oscillatory behavior of Krylov complexity due to the discrete spectra. On the other hand, in the case of continuous spectra (without upper bound), the Krylov complexity shows an exponential increase. Such examples are field theories in non-compact space and the IP model with non-zero mass at non-zero temperature. 

However, the continuous spectrum is not sufficient to show the exponential growth of Krylov complexity. In the IP model example, there is a case where the Krylov complexity shows a linear increase, although it is a continuous spectrum. This is the massless adjoint case, in which case the spectrum shows the Wigner-type behavior with a single Wigner semicircle. On the other hand, when the spectrum can be approximated as an infinite series of Wigner semicircles, the Krylov complexity shows an exponential increase. Therefore, even in the continuous spectrum, the Krylov complexity does not show an exponential increase if there is a clear upper or lower bound in the spectrum of a two-point function. In the IP model example, with increasing temperature, a spectrum consisting of a series of Wigner semicircles ``melds"  to form a smooth continuous spectrum. This does not happen with a single Wigner semicircle for the massless case.

From the above, we consider the case where there is no upper or lower bound in the continuous spectrum of a two-point function as a condition for the Krylov complexity to increase exponentially, but as originally argued in \cite{Parker:2018yvk}, the structure of high-energy tail in the spectrum also needs to be discussed further. In this regard, a clear claim can be made since the high-energy tail of the spectrum determines an asymptotic behavior of the Lanczos coefficients at large $n$, which in turn affects the late-time behavior of Krylov complexity. As studied in \cite{Parker:2018yvk, Barbon:2019wsy}, for the asymptotic growth of Lanczos coefficients to be proportional to $n$ and for the Krylov complexity to increase exponentially, the tail of the spectrum must fall exponentially, which is ``slowly" compared to that of Gaussian types.

Based on these experiences, we propose the following conditions for the spectrum in the case of an exponential increase in Krylov complexity. 
\begin{center}
{A. The spectrum of $\mathcal{O}$ must be continuous rather than discrete.

B. There are no upper and lower bounds in the spectrum, \\and the high-energy tail of the spectrum exponentially falls.}
\end{center}
As explained in Appendix \ref{App0}, the spectrum in these conditions is not an energy spectrum of Hamiltonian but rather the spectrum of a two-point function of $\mathcal{O}$.\footnote{If we consider a retarded two-point function with a step function $\theta(t)$, its spectrum has a non-zero imaginary part. In such a case, as demonstrated in \cite{Iizuka:2023pov}, one can construct a spectrum of a two-point function without the step function from a real part of the spectrum, spectral density, of the retarded two-point function. The Krylov complexity can then be computed from the constructed spectrum.}

Practically, the exponential growth of Krylov complexity is measured over a finite time range. One such time scale is the inverse temperature $\beta=1/T$ of quantum systems at finite temperature $T$. If energy intervals in the discrete spectrum are close enough to each other, the spectrum can be regarded as a continuous spectrum for the measurement of Krylov complexity over the finite time range. For the exponential growth at much later times, the spectrum must be even closer to a continuous spectrum. For example, the Krylov complexity of a free massless scalar theory on a sphere for small $R$ initially follows the exponential growth of Krylov complexity in the flat space limit $R\gg \beta$, but its growth stops at a finite value \cite{Avdoshkin:2022xuw}. As $R$ increases, the discrete spectrum at finite $R$ becomes closer to the continuous spectrum at $R \gg \beta$, and the peak value of Krylov complexity increases.

The exponential fall in condition B means that the spectrum decays exponentially at large $\vert\omega\vert$ as
\begin{align}
G(\omega)\sim e^{-\kappa\vert\omega\vert} \;\; (\vert\omega\vert\to\infty),
\end{align}
where $G(\omega)$ is the spectrum of a two-point function of $\mathcal{O}$, and $\kappa$ is a constant. More precisely, a log correction such as $e^{-\kappa\vert\omega\vert\log\vert\omega\vert}$ can be included in a one-dimensional spin chain with a finite range interaction.
Note that the slowest decay of $G(\omega)$ for a lattice system of fermions with a local Hamiltonian is bounded as \cite{Abanin:2015} 
\begin{align}\label{boundG}
G(\omega)\le C e^{-\kappa\vert\omega\vert},
\end{align}
where $C$ is a constant. Here, for the bound of $G(\omega)$ (\ref{boundG}), the operator $\mathcal{O}$ and local interactions in the Hamiltonian should be $k$-local, and their norms should be finite.

As a further concrete example, the two-dimensional holographic CFT calculation explicitly shows that the behavior of Krylov complexity changes significantly when the spectrum switches from discrete to continuous, where this change of the spectrum in the two-dimensional holographic CFTs indicates a confinement/deconfinement transition of large $N$ theories. For the above reasons, we propose that the Krylov complexity can be an order parameter for rather confinement-like phenomena. \\

\noindent{\bf Example 3: An infinite number of free scalars with various masses in compact space}\\
\noindent Taking the above as a general story of Krylov complexity, in the following we will discuss how the Krylov complexity works as an order parameter in the specific case, such as holographic QCD treated in Section \ref{sec3}, by considering a model with a very similar spectrum.\\

\noindent First, compactification must be applied in the spatial direction. This can be understood from the dispersion relation $E^2=M_m^2+\vec{k}^2$. In the confinement phase, various mesons appear as color singlets, where their mass spectrum of $M_m$ is discrete. However, if $\vec{k}$ is a continuous quantity, then $E$ will be continuous regardless of the discreteness of $M_m$. Krylov complexity is sensitive to the discreteness of $E$, but cannot distinguish between the discreteness of $M_m$ and $\vec{k}$. Therefore, if $\vec{k}$ is continuous, $K(t)$ cannot capture the change in the discreteness of $M_m$ due to the phase transition. Therefore, the spatial directions must be compactified so that a KK tower is sufficiently discrete.\\

In the following, we consider a model consisting of a set of many free scalar fields with different masses in compact space. Specifically, we impose a periodic boundary condition $x = x+L$ from the earlier discussion. This is a simplified version of the spectrum that appears in holographic QCD as treated in Section \ref{sec3}, where various masses of the scalars correspond to various masses of the mesons. First, consider a model consisting of a single free scalar field in $(1+1)$ dimensions. The Euclidean action is
\begin{align}
\label{Ex3-1}
S = \int_{\mathbb{S}^1 \times \mathbb{S}^1} d^2x \left( \frac{1}{2}\partial^{\mu} \phi \partial_{\mu} \phi + \frac{1}{2}m^2 \phi^2 \right).
\end{align}
By treating the momentum in the spatial direction as a KK tower due to the boundary condition, this theory becomes just a sum of harmonic oscillators that depend on $m$ and the discrete momentum. The Wightman inner product correlator for inverse temperature $\beta$ of the harmonic oscillator $H = \frac{p^2}{2m} + \frac{1}{2}m\omega_0^2 x^2$ is as given in Appendix \ref{App0},
\begin{align}
G(t) = \frac{\hbar \cos [\hbar\omega_0 t]}{2m\omega_0 \sinh[\hbar\omega_0\beta/2]} .
\end{align}
To map the Hamiltonian of the harmonic oscillator to \eqref{Ex3-1}, simply replace $m, \hbar \to 1$ and $\omega_0 \to \sqrt{m^2 + \left(\frac{2\pi \ell}{L}\right)^2},\ (\ell=0,\pm1,\pm2,\cdots)$. Here, the contribution of the 
KK tower is considered as a sum with respect to $\ell$. As a result, we obtain
\begin{align}
G(t) = \sum_{\ell=-\infty}^{\infty} \frac{\cos [t\sqrt{m^2 + \left(\frac{2\pi \ell}{L}\right)^2 }]}{2\sqrt{m^2 + \left(\frac{2\pi \ell}{L}\right)^2}\sinh\left[\frac{\beta}{2}\sqrt{m^2 + \left(\frac{2\pi \ell}{L}\right)^2}\right]} .
\end{align}
Now, for simplicity let us take $\beta / L \gg1$. As $\sinh$ with nonzero $\ell$ in the denominator increases exponentially, nonzero $\ell$ terms can be ignored, so that it can be approximated that only $\ell=0$ contributes by taking large $\beta / L$. In this case, the Krylov complexity just oscillates
\begin{align}\label{KCHO}
G(t) \propto \cos (mt)\ \to\ K(t) = \sin^2 (mt).
\end{align}
Whatever the value of $m$, $K(t)$ for large enough $\beta / L$ oscillates and does not grow exponentially.
The situation does not change when this model is in (3+1)-dimension. Specifically, if a periodic boundary condition is imposed on each of the three spatial directions, the contribution of the KK tower corresponds to 
\begin{align}
\omega_0 \to  \sqrt{m^2 + \left(\frac{2\pi}{L}\right)^2 (\ell_x^2+\ell_y^2+\ell_z^2)}\,. 
\end{align}
Under $\beta/L \gg 1$, only $\ell_x=\ell_y=\ell_z=0$ term contributes as in the (1+1)-dimensional case, so $G(t)$ shows just an oscillation and the Krylov complexity shows no exponential growth.\\

In the above, we saw that, for large $\beta/L$, the Krylov complexity does not increase exponentially. Next, we consider a model that mimics the spectrum as treated in Section \ref{sec3}. The model is a (1+1)-dimensional system consisting of many free scalar fields $\{ \phi_n(t,x) \}$ whose masses are displaced by $\delta m $, and the smallest mass is $m$. We impose the simple periodic boundary condition $x = x+L$. The Euclidean action is
\begin{align}
\label{ex3model}
S = \int_{\mathbb{S}^1 \times \mathbb{S}^1} d^2x \sum_{n=0}^{\infty}\left( \frac{1}{2}\partial^{\mu} \phi_n \partial_{\mu} \phi_n + \frac{1}{2}m_n^2 \phi_n^2 \right),\ \ m_n = m +n \delta m.
\end{align}
The system has four dimensionful parameters $m,\delta m, L, \beta$. Therefore, we can adopt $\beta \delta m, \beta/L$ and $\beta m$ as dimensionless parameters. Note here that the model of \eqref{Ex3-1} corresponds to $\beta \delta m \gg 1$, where nonzero $n$ terms can be ignored. Also, by analogy with Section \ref{sec3}, the confinement phase corresponds to the case where $\beta\delta m\gtrsim 1$ and the deconfinement phase corresponds to the case where $\beta \delta m \ll 1$.

We consider the Wightman inner product correlator of a composite operator $\mathcal{O}=\sum_{n=0}^\infty \phi_n$. In QCD, we consider a color singlet operator such as $\mathcal{O}=\Tr[F_{\mu\nu}F^{\mu\nu}]$, and thus $\phi_n$ corresponds to various glueballs with different masses, where $n$ is an index for ``radial" Regge trajectories. Since the system is the free theory, the two-point function of $\mathcal{O}$ is a sum of the two-point function of $\phi_n$. In fact, in large $N$ theory, the glueballs can be treated as free fields.
Thus, by keeping in mind the application to QCD, we study the composite operator $\mathcal{O}=\sum_{n=0}^\infty \phi_n$ in our model \eqref{ex3model}. 
The Wightman inner product correlator of $\mathcal{O}$ of the model \eqref{ex3model} is obtained immediately, as in the previous example, 
\begin{align}\label{FreeGtExact}
G(t) = \sum_{n=0}^\infty\sum_{\ell=-\infty}^{\infty} \frac{\cos [t\sqrt{(m+n\delta m)^2 + \left(\frac{2\pi \ell}{L}\right)^2 }]}{2\sqrt{(m+n\delta m)^2 + \left(\frac{2\pi \ell}{L}\right)^2}\sinh\left[\frac{\beta}{2}\sqrt{(m+n\delta m)^2 + \left(\frac{2\pi \ell}{L}\right)^2}\right]} .
\end{align}

In fact, the two-point functions in holographic QCD as treated in Section \ref{sec3} have a similar sum structure with respect to $n$ and $\ell$. In particular, the sum for $\ell$ corresponds to the KK tower for a compact space in which the QCD lives, and the sum for $n$ corresponds to the KK tower for emergent radial direction.

As before, consider the case of $\beta/L \gg 1$ so that only $\ell=0$ contributes. Also, focusing on the region $\beta m \gg 1$ to obtain an analytic expression, the summation with respect to $n$ can be performed
\begin{align}
G(t) &\sim \sum_{n=0}^{\infty} \frac{\cos [t(m+n\delta m)]}{(m+n\delta m)\exp\left[\frac{\beta}{2}(m+n\delta m)\right]} \notag\\
&= \frac{1}{2\delta m}\left( e^{-i mt-\frac{m\beta}{2}} \Phi(e^{-i \delta m t-\frac{\delta m\beta}{2}},1,m/\delta m) + e^{i mt-\frac{m\beta}{2}} \Phi(e^{i \delta m t-\frac{\delta m\beta}{2}},1,m/\delta m)\right),\label{GtManyScalar}
\end{align}
where $\Phi(z,s,a)$ is Hurwitz-Lerch transcendental function defined by
\begin{align}
\Phi(z,s,a):=\sum_{n=0}^\infty z^n (n+a)^{-s}.
\end{align}
 Using this two-point function, the Lanczos coefficient and the Krylov complexity can be obtained numerically. In the following, we perform numerical computations for $\beta=1$ and $m=10$. In such numerical computations, $1/L=5$ is large enough to approximate (\ref{FreeGtExact}) by (\ref{GtManyScalar}).
 
\begin{figure}
\centering
  \begin{subfigure}[b]{0.6\textwidth}
         \centering
         \includegraphics[width=\textwidth]{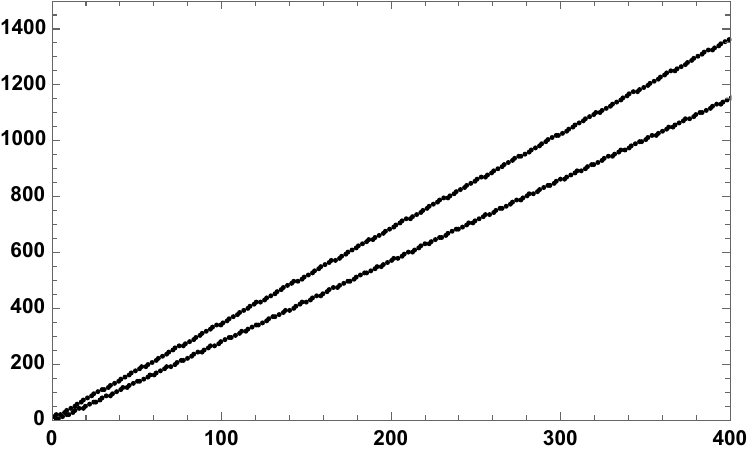}
       \put(5,5){$n$}
    \put(-255,160){$b_n$}
       \caption{Lanczos coefficient $b_n$ up to $n=400$}\label{fig:bnbeta1m10dm5}
     \end{subfigure}
     \begin{subfigure}[b]{0.6\textwidth}
         \centering
         \includegraphics[width=\textwidth]{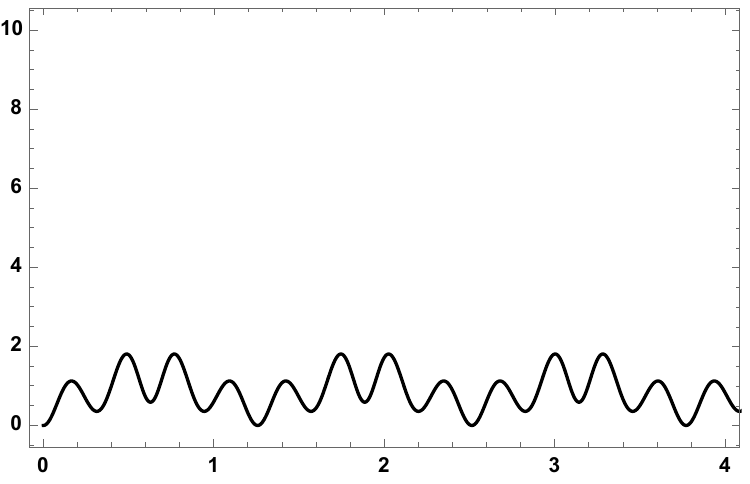}
       \put(5,5){$t$}
    \put(-255,170){$K(t)$}
       \caption{Krylov complexity $K(t)$ that is computed from $b_n$ up to $n=400$}\label{fig:KCbeta1m10dm5}
     \end{subfigure}
               \caption{Lanczos coefficient $b_n$ and Krylov complexity $K(t)$ of $G(t)$ (\ref{FreeGtExact}) for $\beta=1$, $1/L=5$, $m=10$, $\delta m=5$.   }
        \label{fig:beta1m10dm5}
\end{figure}

Figure \ref{fig:beta1m10dm5} shows the Lanczos coefficient $b_n$ and the Krylov complexity $K(t)$ computed numerically from $G(t)$ (\ref{FreeGtExact}) for $\beta=1$, $1/L=5$, $m=10$, $\delta m=5$.
We can see that $b_n$ obeys the two-slopes behavior and $K(t)$ oscillates and does not grow, which are characteristic behaviors for the discrete spectrum \cite{Avdoshkin:2022xuw} due to $\beta \delta m=5 \sim \order{1}$.

\begin{figure}
\centering
  \begin{subfigure}[b]{0.6\textwidth}
         \centering
         \includegraphics[width=\textwidth]{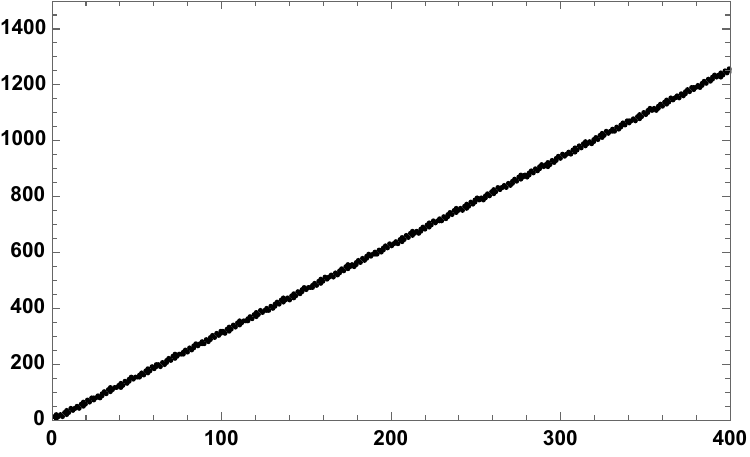}
       \put(5,5){$n$}
    \put(-255,160){$b_n$}
       \caption{Lanczos coefficient $b_n$ up to $n=400$}\label{fig:bnbeta1m10dm1100}
     \end{subfigure}
     \begin{subfigure}[b]{0.6\textwidth}
         \centering
         \includegraphics[width=\textwidth]{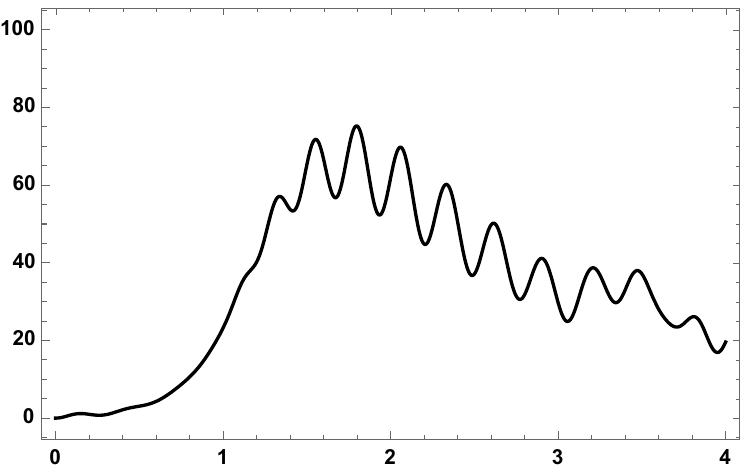}
       \put(5,5){$t$}
    \put(-255,170){$K(t)$}
       \caption{Krylov complexity $K(t)$ that is computed from $b_n$ up to $n=400$}\label{fig:KCbeta1m10dm1100}
     \end{subfigure}
          \begin{subfigure}[b]{0.6\textwidth}
         \centering
         \includegraphics[width=\textwidth]{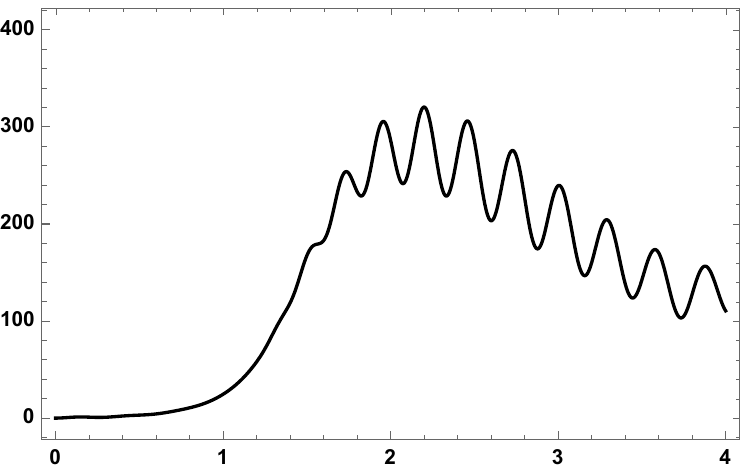}
       \put(5,5){$t$}
    \put(-255,170){$K(t)$}
       \caption{Krylov complexity $K(t)$ that is computed from the extrapolated $b_n$ from $n=400$ up to $n=2000$}\label{fig:KCbeta1m10dm1100ex}
     \end{subfigure}
               \caption{Lanczos coefficient $b_n$ and Krylov complexity $K(t)$ of $G(t)$ (\ref{FreeGtExact}) for $\beta=1$, $1/L=5$, $m=10$, $\delta m=1/100$. The Krylov complexity $K(t)$ initially grows, but the increase stops since we only use a finite number of $b_n$ to compute $K(t)$ numerically.  }
        \label{fig:beta1m10dm1100}
\end{figure}
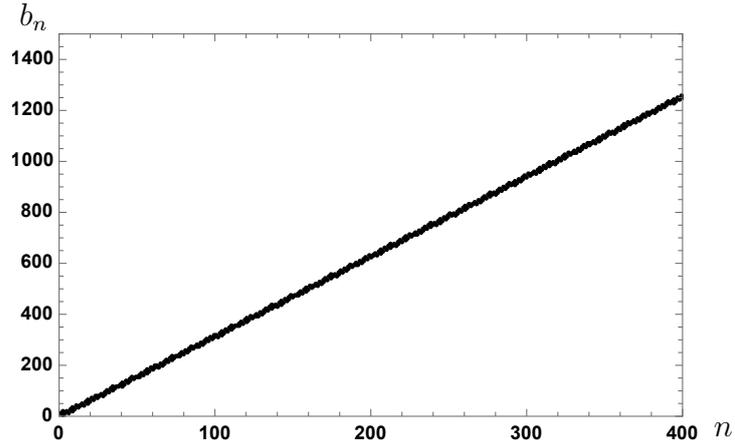
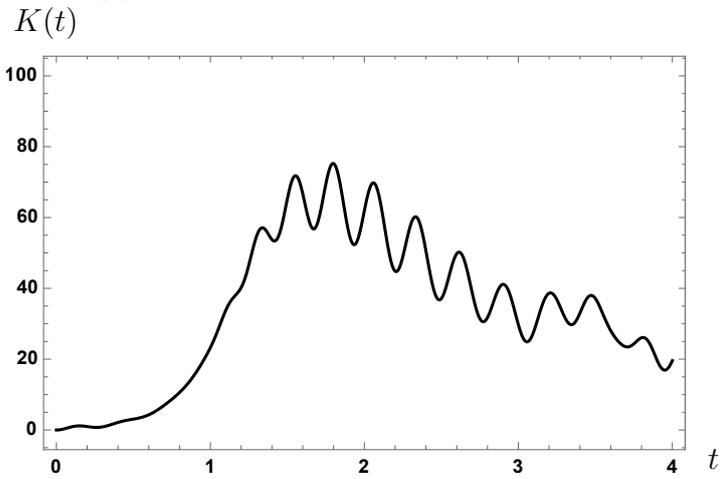
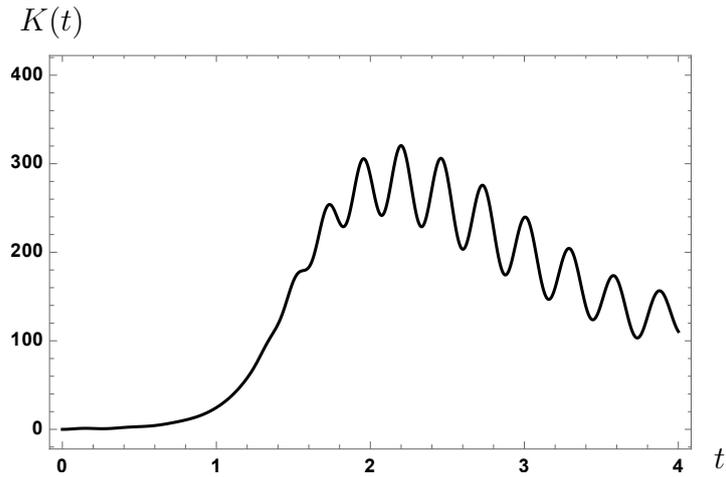

Figure \ref{fig:beta1m10dm1100} shows the Lanczos coefficient $b_n$ and the Krylov complexity $K(t)$ computed numerically from $G(t)$ (\ref{FreeGtExact}) for $\beta=1$, $1/L=5$, $m=10$, $\delta m=1/100$. In Figure \ref{fig:bnbeta1m10dm1100}, $b_n$ does not show the two-slope behavior because the spectrum is close to continuous due to small $\beta \delta m=1/100  \ll \order{1}$.
Figure \ref{fig:KCbeta1m10dm1100} shows $K(t)$ computed from $b_n$ up to $n=400$, where $K(t)$ initially grows rapidly compared to $K(t)$ in Figure \ref{fig:KCbeta1m10dm5}, which implies the exponential growth due to small $\beta \delta m=1/100 $.
However, the increase stops around $K(t)\sim80$ since we only use $b_n$ up to $n=400$ to compute $K(t)$. Due to the infinite sum in (\ref{FreeGtExact}), $b_n$ is nonzero even at $n\to\infty$. However, in numerical computations, we can only use a finite number of $b_n$. If we use $b_n$ with an even larger $n$ to compute $K(t$), $K(t)$ is expected to grow even more. To confirm this expectation, we extrapolate $b_n$ in Figure  \ref{fig:bnbeta1m10dm1100} up to $n=2000$ and compute $K(t)$ from the extrapolated $b_n$.\footnote{Due to $m=10$, $b_n$ is divided into two families for even $n$ and for odd $n$ \cite{Avdoshkin:2022xuw, Camargo:2022rnt}. Thus, we extrapolate $b_n$ separately for even $n$ and odd $n$.} As shown in Figure \ref{fig:KCbeta1m10dm1100ex}, $K(t)$ further grows up to $K(t)\sim300$.\\

\clearpage

\noindent{\bf Proposal}\\
In example 3, it is important to note that the interval $\delta m$ of the mass spectrum, not the smallest mass $m$, affects the continuity of the spectrum associated with changes in the behavior of Krylov complexity. In this example, the spectrum can be approximated as continuous if either $\beta \delta m \ll 1$ or $\beta / L \ll 1$ is achieved. The spectrum of two-point functions in holographic QCD has a similar property, where there are two KK towers for extra dimensions and for a compact space in which the QCD lives. Since we are interested in measuring a confinement/deconfinement phase transition, we take the temperature near the phase transition, such as the QCD scale $\Lambda_{\rm QCD}$.
As mentioned at the beginning of example 3, to examine (de)confinement due to the mass spectrum, the compactification has to be done properly. If the KK tower associated with the compactification can be approximated as continuous, as in example 1, it is not possible to properly examine (de)confinement. Therefore, we should take $\beta / L \gtrsim 1$, such that the KK tower associated with the compactification can be regarded as discontinuous. Then, the continuity of the spectrum is determined by $\beta\delta m$, where $\beta\delta m\gtrsim 1$ is the confinement phase, and $\beta\delta m \ll 1$ is the deconfinement phase.

From the above, we propose that the Krylov complexity can be used as an order parameter of a confinement/deconfinement phase transition in large $N$ field theories by the following prescription. \\

{\bf I. Take temperature near the confinement/deconfinement phase transition such as the QCD scale $\beta \sim \Lambda_{\rm QCD}^{-1}$.

II. Next, compactify space sufficiently $\beta / L \gtrsim 1$ for discrete momentum.

III. Then, the continuity of the spectrum is determined by $\beta \delta m$ in the mass spectrum, 
and the Krylov complexity works as an order parameter.}
\\

In the next section, we will confirm this in holographic QCD.

\section{Krylov complexity for holographic Yang-Mills theories}
\label{sec3}
In this section, we give further examples of the Krylov complexity as an order parameter by studying $SU(N)$ Yang-Mills theories in the large $N$ limit via holography. We analyze their spectrum by using the holographic method and evaluate their Krylov complexity as an order parameter of the Hawking-Page transition in the bulk that corresponds to a confinement/deconfinement transition in the large $N$ field theory side.
Of course, as we can learn from the free scalar example, proper compactification is necessary.

\subsection{Krylov complexity for $\mathcal{N}=4$ Super Yang-Mills theory}
In this subsection, we first consider the $\mathcal{N}=4$ Super Yang-Mills theory in the large $N$ limit from holography \cite{Maldacena:1997re}, and then look at its behavior on the (de)confinement of the Krylov complexity. Of course, the $\mathcal{N}=4$ SYM theory at finite temperature on $\mathbb{R}^3$ has only one dimensionful parameter: temperature, so no phase transition occurs even in the large $N$ limit. However, in the large $N$ $\mathcal{N}=4$ SYM theory at finite temperature on $\mathbb{S}^3$, there exists a phase transition, and the entropy of the system changes from $\order{1}$ to $\order{N^2}$ \cite{Witten:1998qj,Witten:1998zw}. In the bulk description, this is the Hawking-Page transition \cite{Hawking:1983tob} between Thermal-AdS and Schwatzchild AdS black hole in global coordinates.
Specifically, we calculate the spectrum of the bulk scalar and glueball on the background in this configuration. 

\subsubsection{Thermal AdS}
First, consider the AdS$_5$ space-time in global coordinates. We consider a situation in which a scalar field $\phi$ with mass $m$ propagates in this bulk and check the mass gap. Starting from
\begin{align}\label{thermalAdSmetric}
ds^2&=-(r^2+1)dt^2 + \frac{dr^2}{r^2+1} + r^2 d\Omega_3^2,
\end{align}
where this metric is given by global coordinates. In the Euclidean signature, the AdS boundary is thermal circle $\times$ compact space (Sphere), $\mathbb{S}^1 \times \mathbb{S}^3$. On this background, the field equation for $\phi$ is
\begin{align}
\frac{1}{\sqrt{-g}}\partial_{\mu}(\sqrt{-g}g^{\mu \nu}\partial_{\nu}\phi)-m^2 \phi=0,
\end{align}
and we decompose $\phi=f(r)e^{-i \omega t}Y_{\ell\vec{m}}(\Omega)$, then 
\begin{align}
\frac{1}{r^3}\partial_r [r^3 (r^2+1) \partial_r f(r)] +\left( \frac{\omega^2}{r^2+1} -\frac{\ell(\ell+2)}{r^2}-m^2\right) f(r)=0.
\end{align}
Here, $Y_{\ell\vec{m}}(\Omega)$ is a spherical harmonic on $\mathbb{S}^3$ with an eigenvalue $\ell(\ell+2)$.

\noindent{\bf \underline{Near the boundary, $r \to \infty$}}\\
Near the boundary, this equation becomes
\begin{align}
\frac{1}{r^3}\partial_r [r^5 \partial_r f(r)]-m^2f(r)=0
\end{align}
This equation has the following solution
\begin{align}
f(r)&=c_1 r^{-\Delta}+c_2 r^{\Delta-4},\\
\Delta&=2+\frac{1}{2}\sqrt{4^2+4m^2}.
\end{align}
Furthermore, given the extreme limit to the boundary, it is understood that $c_2=0$ is necessary to extract normalizable mode and that $f(r)$ behaves as $r^{-\Delta}$ near the boundary as a boundary condition.\\

\noindent{\bf \underline{Near the center, $r \to 0$}}\\
On the other hand, near the center of AdS$_5$, $r=0$, the equation of motion becomes
\begin{align}
\partial_r^2 f(r)+\frac{3}{r} \partial_r f(r) -\frac{\ell(\ell+2)}{r^2}f(r)=0,
\end{align}
and a solution is 
\begin{align}
f(U)=c_{c1} r^{-1-\sqrt{1+\ell(\ell+2)}}+c_{c2}r^{-1+\sqrt{1+\ell(\ell+2)}}
\end{align}
Smoothness at $r=0$ requires $c_{c1}=0$.\\

It is known that the full equation can be solved in terms of Hypergeometric functions, and the spectrum of $\omega$ is quantized under the constraint that the boundary conditions of both the AdS boundary and the center of the bulk are satisfied. As a result, the spectrum of $\omega$ is given by the following quantized values
\begin{align}\label{SpectrumThermalAdS}
\omega = \Delta + \ell + 2n\ \ (\ell=0,1,2,\cdots,\ n=0,1,2,\cdots).
\end{align}
This sum structure with respect to $n$ and $\ell$ is very similar to example 3 in Section \ref{sec2}, where the length scale in (\ref{SpectrumThermalAdS}) is measured by the AdS scale $R=1$, and $\Delta$ corresponds to the smallest mass in example 3.
Since this is a discrete spectrum, the Krylov complexity exhibits oscillation.

\subsubsection{BH case}
The analysis of this subsection follows \cite{Festuccia:2005pi}. The $\mathcal{N}=4$ SYM theory in the large $N$ limit on $\mathbb{S}^3$ at high temperature is dual to the following Schwarzschild black hole bulk geometry 
\begin{align}\label{AdS5BH}
ds^2&=-f(r)dt^2 + \frac{dr^2}{f(r)} + r^2 d\Omega_3^2,\\
&f(r)=r^2+1-\frac{\mu}{r^2}=\frac{1}{r^2}(r^2-r_0^2)(r^2+1+r_0^2),
\end{align}
where $\mu$ is proportional to the mass of the black hole and $r_0$ is the radius of the event horizon. If we take $r \to \infty$, this spacetime approaches to global AdS$_5$. On this background, we decompose $\phi=e^{-i\omega t}r^{-\frac{3}{2}}\psi$, then the equation of motion is given by
\begin{align}
\label{AdS5Seq}
&(-\partial_z^2 + V(z) -\omega^2)\psi = 0,\\
&V(z)=f(r)\left[ \frac{3}{4r^2}+\nu^2-\frac{1}{4}+\frac{9\mu}{4r^4} \right],
\end{align}
where $z$ is a tortoise coordinate
\begin{align}
\label{tortoiseSAdS}
z=\int^{\infty}_r \frac{dr}{f(r)},
\end{align}
and $\nu$ is defined by
\begin{align}
\Delta=2+\nu,\ \ \nu=\sqrt{4+m^2}.
\end{align}

At the horizon, the potential becomes $V(z) \to 0$ therefore we can normalize it to be real and thus  \cite{Festuccia:2005pi}
\begin{align}
\label{bconaround0SAdS}
\psi(z) = e^{i \omega z - i \delta} +  e^{-i \omega z + i \delta},
\end{align}
where $\delta$ is a phase shift.
Near the horizon, we have
\begin{align}
z = \int_r^\infty \frac{dr}{f(r)} \propto \int_r^\infty \frac{dr}{r-r_0}  =- \log (r - r_0)  + 
\mbox{const.}\,, 
\end{align}
thus $z$ decreases as we increase $r$. Therefore  $e^{- i \omega z - i \delta}$ is an out-going mode, and $ e^{+ i \omega z - i \delta} $ is an in-going mode. 
For the retarded Green's function, we keep only the ingoing mode, on the other hand, for the advanced Green's function, we keep only the out-going mode. As \cite{Festuccia:2005pi}, we consider the Wightman Green function and thus kept both ingoing and outgoing modes. \\

From \eqref{AdS5Seq}, the spectrum in the high-energy region, which is necessary for the analysis of Krylov complexity, can be derived by the WKB method.
We turn our attention to the following mass regions
\begin{align}
\omega = \nu u,\ \ \nu \gg 1.
\end{align}
By setting $\psi(r)=e^{\nu S}$, we have
\begin{align}\label{WKBequation}
-\nu^2 (\partial_z S)^2 - \nu \partial_z^2 S + \nu^2 f(z) = \nu^2 u^2 + O(\nu^0).
\end{align}
For the WKB method, expanding as $S=S^{(0)}+\frac{1}{\nu}S^{(1)}+\cdots$, 
we obtain
\begin{align}
S^{(0)}  &= \int dz \sqrt{f - u^2} = -\int_{r_c}^{r} dr' \kappa(r') \,, \\
&\mbox{where} \quad \kappa=\frac{1}{f(r)} \sqrt{f(r)-u^2} \,,\\
S^{(1)}&= \log \frac{1}{\left({ V_0 - u^2} \right)^{1/4}},
\end{align}
from (\ref{WKBequation}) in the leading order at $\order{\nu^2}$ and $\order{\nu^1}$.
Here we used \eqref{tortoiseSAdS} and 
\begin{align}
f(r_c) = u^2 \,,
\end{align} 
at the turning point $r = r_c$. 
Therefore under the WKB approximation, the approximate solution becomes
\begin{align}
\psi^{(wkb)}(r)= \frac{1}{\left({ f(r) - u^2} \right)^{1/4}} e^{\nu \mathcal{Z}} \left(1+\order{\nu^{-1}} \right),\ \ \ \mathcal{Z}=-\int_{r_c}^{r} dr' \kappa(r') \,.
\end{align}
Near the horizon $f(r) \to 0$, the boundary condition eq.~\eqref{bconaround0SAdS} determines the relative normalization factor as follows.
\begin{align}
\psi^{(wkb)}(r)=\frac{1}{\sqrt{u}}\psi(r).
\end{align}

Now, the bulk correlator is given by
\begin{align}
\mathcal{G}_+ \sim \frac{1}{\omega}(r r')^{-\frac{3}{2}} \psi(r)\psi(r').
\end{align}
The normalizable mode asymptotically approaches $\sim r^{-\Delta}$. From this, the relationship between boundary correlation function $G_+$ and bulk correlator $\mathcal{G}_+$ is
\begin{align}
G_+ &\sim \lim_{r,r'\to\infty}\ (2 \nu r^{\Delta})(2\nu r'^{\Delta}) \mathcal{G}_+(r,r')\\
& \sim \lim_{r,r'\to\infty}\ 2\nu r^{\Delta}r'^{\Delta} (r r')^{-\frac{3}{2}}\frac{1}{ \left(f(r) - u^2 \right)^{1/4} } e^{\nu \mathcal{Z}(r)}\frac{1}{ \left(f(r') - u^2 \right)^{1/4} } e^{\nu \mathcal{Z}(r')}.
\end{align}
Now, $\left(f(r) - u^2 \right)^{1/4} \to r^{1/2}$ at large $r$,  then
\begin{align}
G_+& \sim \lim_{r,r'\to\infty}\ 2\nu r^{\Delta}r'^{\Delta} (r r')^{-\frac{3}{2}}\frac{1}{r^{1/2} } e^{\nu \mathcal{Z}(r)}\frac{1}{ r'^{1/2}} e^{\nu \mathcal{Z}(r')}\\
&\sim \lim_{r,r'\to\infty}\ 2\nu r^{\nu}r'^{\nu}  e^{\nu \mathcal{Z}}e^{\nu \mathcal{Z}} \sim \lim_{r,r'\to\infty}\ 2\nu e^{2 \nu (\log r + \mathcal{Z}(r))}.
\end{align}
The argument of the exponent is 
\begin{align}
\log r+\mathcal{Z}(r)&=\log r-\int_{r_c}^{r} dr' \frac{1}{f(r)} \sqrt{f(r)-u^2}.
\end{align}

To evaluate the Lanczos coefficient at large $n$, we need to evaluate the Green function at large $\omega$. The turning point $r_c$ becomes large at large $\omega$, and therefore in such a large $r_c$, $f(r) \approx r^2$ and $r_c \approx u$, then we have 
\begin{align}
\lim_{r \to \infty} \left[ \log r+\mathcal{Z} \right]
&\simeq \lim_{r \to \infty} \left[ \log r - \int_{r_c=u}^{r} dr'  \frac{\sqrt{r^2-u^2}}{r^2}  \right]\\
&=  \lim_{U \to \infty} \left[ \log r - \int_1^{u^{-1}r} dx \frac{\sqrt{x^2-1}}{ x^2}  \right] \\
&\simeq 1+\log \frac{r_c}{2}.
\end{align}
Thus the boundary correlation function $G_+(\omega)$ in the large $\omega$ limit becomes 
\begin{align}
\label{Gplusourcase}
G_+(\omega)\propto \omega^{2\nu}  \quad \mbox{at large $\omega$}.
\end{align}
Therefore the spectrum grows exponentially in $\omega$, where $\nu = m + \order{1/m}$ in large $m$ for WKB approximation.

The Wightman inner product correlator for UV regulation is
\begin{align}
G_{12}(t) = \Tr \left[ e^{-\beta H} {\cal{O}}(t - i\beta/2)  {\cal{O}}(0) \right].
\end{align}
In frequency space, $G_{12}(\omega)$ and $G_+(\omega)$ are related as 
\begin{align}\label{G12w}
G_{12}(\omega) = e^{-\frac{\omega \beta}{2}} G_+(\omega).
\end{align}
This can be understood simply by changing the contour of the $t$ integration in the Fourier transformation.  Then 
\begin{align}\label{asbG12BH}
G_{12}(\omega)\propto \omega^{2 \nu} e^{  -\frac{\omega \beta}{2}}  \to 0 \quad \mbox{at large $\omega$} \,,
\end{align}
goes to zero at large $\omega$, {\it i.e.,} UV regulated.

Next, we use this Green function to evaluate the Krylov complexity. However before that, let us comment on the validity of the WKB approximation. For the WKB approximation, the dimension of the operator $\mathcal{O}$ needs to be large. 
However, the continuum properties of the Green function are essentially determined by the boundary condition on the black hole horizon. Thus, although we use the results of WKB approximation for the Green function to evaluate the Krylov complexity, we expect that the resultant Krylov complexity are not so much dependent on the detail of WKB approximations.

\subsubsection{Krylov complexity of the discrete spectrum}\label{sec:313}
We evaluate the Krylov complexity associated to the two-point function $G_{12}(t)$, which is symmetric with respect to $t$:
\begin{align}
G_{12}(t)&=\Tr \left[ e^{-\beta H} {\cal{O}}(t - i\beta/2)  {\cal{O}}(0) \right]=\text{Tr}[e^{-\beta H/2}e^{i H t} {\cal{O}} e^{-i H t} e^{-\beta H/2}{\cal{O}}]\notag\\
&=\text{Tr}[e^{-\beta H/2}e^{-i H t} {\cal{O}} e^{+i H t} e^{-\beta H/2}{\cal{O}}]=G_{12}(-t).
\end{align}
In frequency space, $G_{12}(\omega)$ is also symmetric with respect to $\omega$:
\begin{align}
G_{12}(\omega):=\int_{-\infty}^{\infty}dt\,e^{i\omega t}G_{12}(t)=\int_{-\infty}^{\infty}dt\,e^{i\omega t}G_{12}(-t)=G_{12}(-\omega).
\end{align}
From the spectrum (\ref{SpectrumThermalAdS}) for thermal AdS and (\ref{G12w}), we consider the following discrete spectrum
\begin{align}\label{G12N4thermalAdS}
G_{12}(\omega)=\frac{1}{N_0}\sum_{\ell=0}\sum_{n=0}e^{-\frac{\vert\omega\vert \beta}{2}}\left[\delta\left(\omega-(\Delta + \ell + 2n)\right)+\delta\left(\omega+(\Delta + \ell + 2n)\right)\right],
\end{align}
where $N_0$ is a normalization constant such that
\begin{align}
\int_{-\infty}^{\infty}\frac{d\omega}{2\pi}G_{12}(\omega)=1.
\end{align}

Figure \ref{fig:KCN4thermalAdS} shows the Lanczos coefficient $b_n$ and the Krylov complexity $K(t)$ computed numerically from the discrete spectrum $G_{12}(\omega)$ (\ref{G12N4thermalAdS}) for $\Delta=10$ and $\beta=2\pi$. 
The length scale that determines the discreteness of (\ref{G12N4thermalAdS}) is the AdS scale $R=1$. Since $\beta/R=2\pi \sim \order{1}$, $b_n$ and $K(t)$ behave similar to those in Figure \ref{fig:beta1m10dm5}.

\newpage

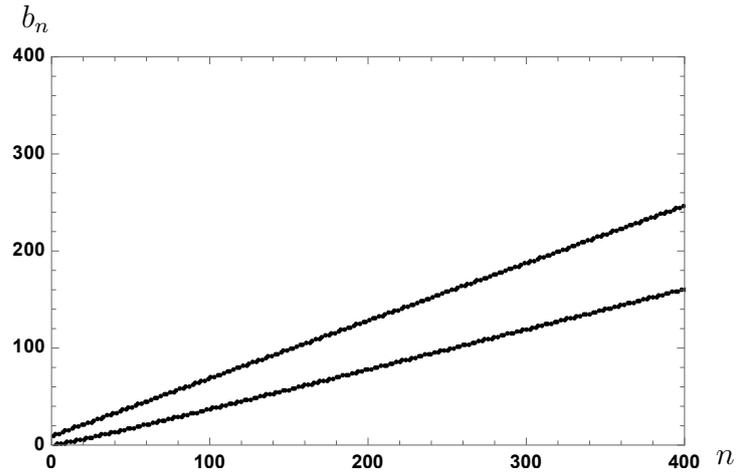
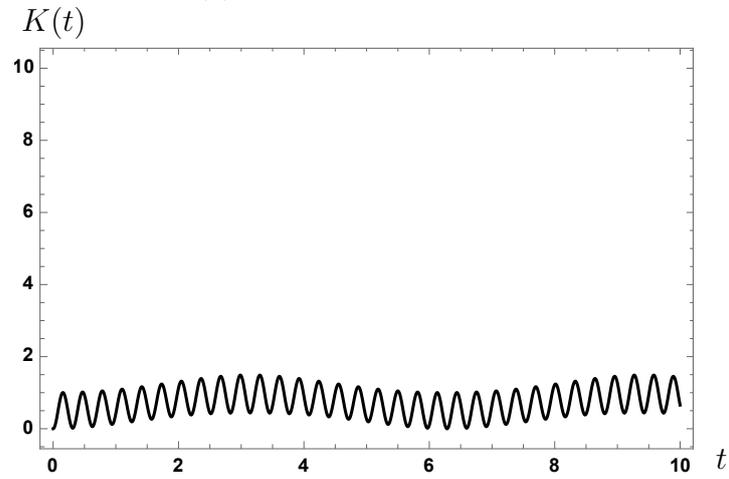
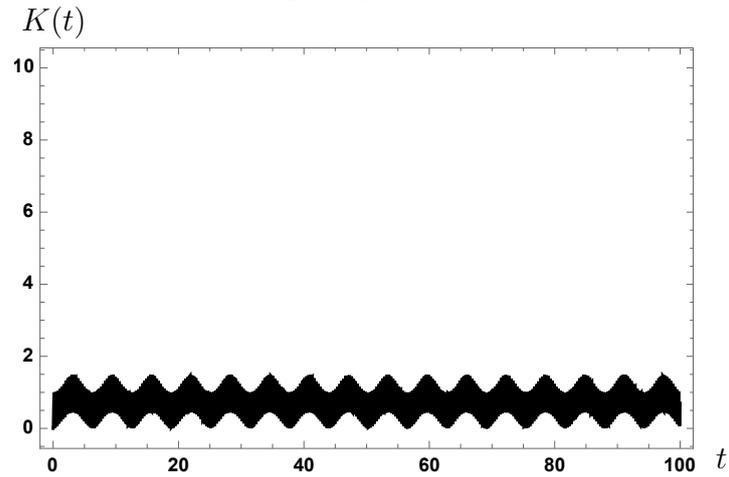
\begin{figure}
\centering
     \begin{subfigure}[b]{0.6\textwidth}
         \centering
         \includegraphics[width=\textwidth]{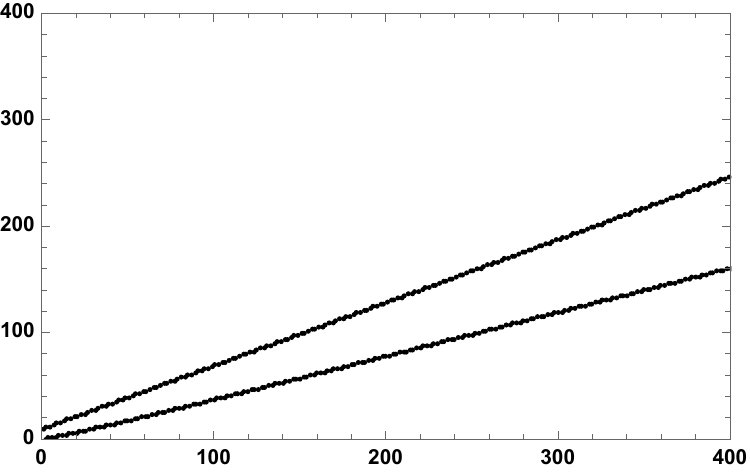}
       \put(5,5){$n$}
    \put(-255,170){$b_n$}
       \caption{Lanczos coefficient $b_n$}\label{fig:LanczosN4thermalAdS}
     \end{subfigure}
           \begin{subfigure}[b]{0.6\textwidth}
         \centering
         \includegraphics[width=\textwidth]{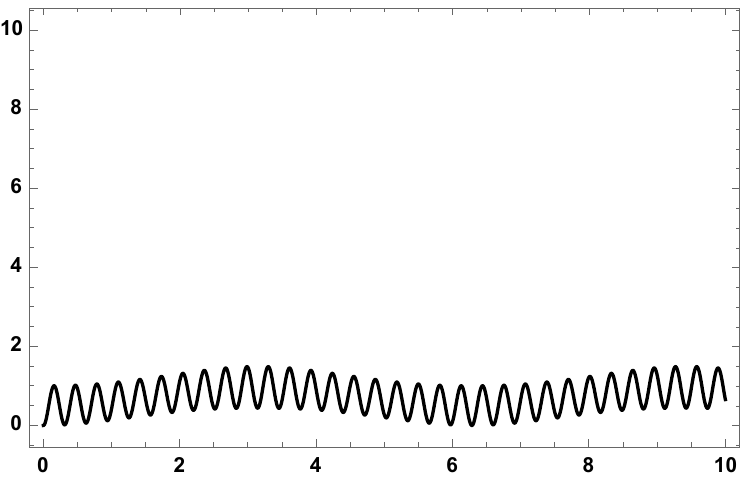}
       \put(5,5){$t$}
    \put(-255,170){$K(t)$}
       \caption{Krylov complexity $K(t)$ in $0\le t\le10$}\label{fig:KCN4thermalAdS}
     \end{subfigure}
                \begin{subfigure}[b]{0.6\textwidth}
         \centering
         \includegraphics[width=\textwidth]{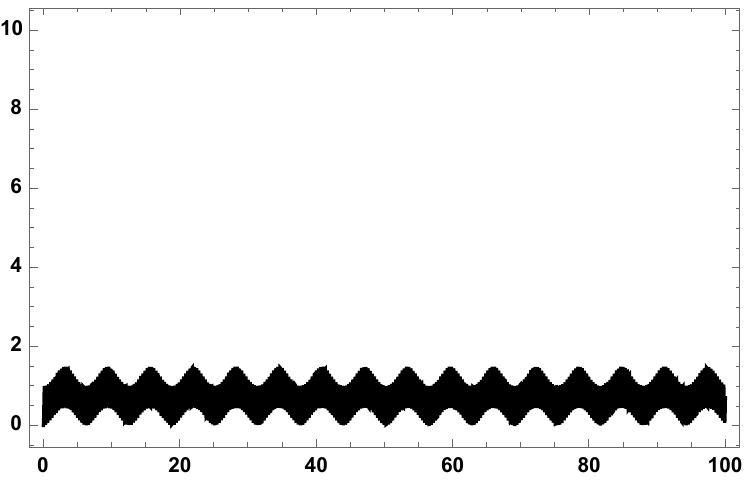}
       \put(5,5){$t$}
    \put(-255,170){$K(t)$}
       \caption{Krylov complexity $K(t)$ in $0\le t\le100$}\label{fig:KCN4thermalAdSLatetime}
     \end{subfigure}
             \caption{Numerical plots of the Lanczos coefficient $b_n$ and the Krylov complexity $K(t)$ of the discrete spectrum $G_{12}(\omega)$ (\ref{G12N4thermalAdS}) for $\Delta=10$ and $\beta=2\pi$.}
        \label{fig:KCN4thermalAdS}
\end{figure}

\clearpage

\subsubsection{Krylov complexity of the continuous spectrum}
\begin{figure}
\centering
  \begin{subfigure}[b]{0.6\textwidth}
         \centering
         \includegraphics[width=\textwidth]{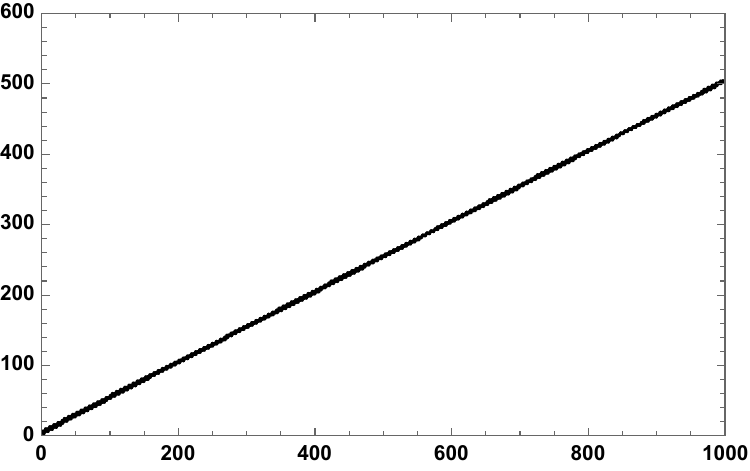}
       \put(5,5){$n$}
    \put(-255,170){$b_n$}
       \caption{Lanczos coefficient $b_n$}\label{fig:LanczosN4BH}
     \end{subfigure}
     \begin{subfigure}[b]{0.6\textwidth}
         \centering
         \includegraphics[width=\textwidth]{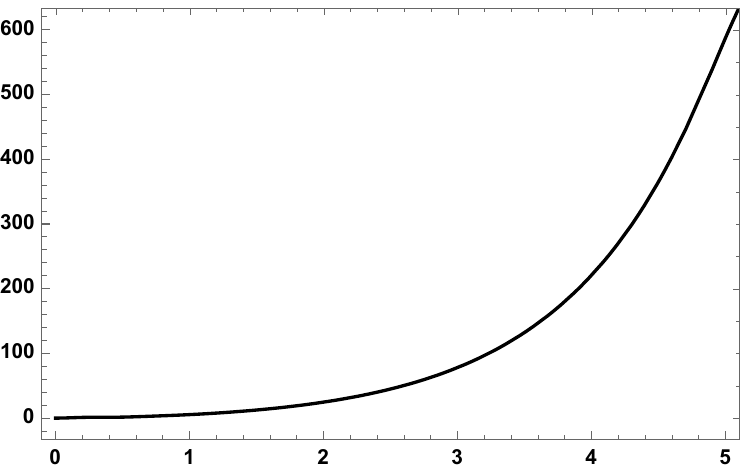}
       \put(5,5){$t$}
    \put(-255,170){$K(t)$}
       \caption{Linear plot $K(t)$ of the Krylov complexity}\label{fig:KCN4BHlinear}
     \end{subfigure}
           \begin{subfigure}[b]{0.6\textwidth}
         \centering
         \includegraphics[width=\textwidth]{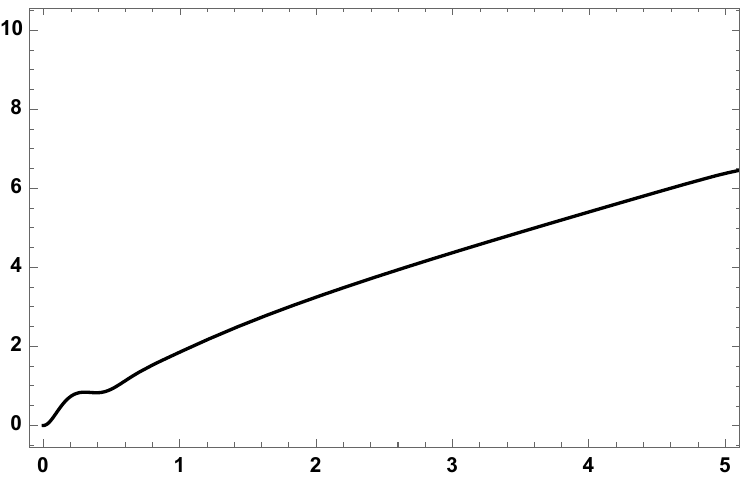}
       \put(5,5){$t$}
    \put(-275,170){$\log[1+K(t)]$}
       \caption{Log plot $\log[1+K(t)]$ of the Krylov complexity}\label{fig:KCN4BHlog}
     \end{subfigure}
             \caption{Numerical plots of the Lanczos coefficient $b_n$ and the Krylov complexity $K(t)$ of the continuous spectrum $G_{12}(\omega)$ (\ref{G12N4BH}) for $\nu=10$ and $\beta=2\pi$. 
             }
        \label{fig:KCN4BH}
\end{figure}
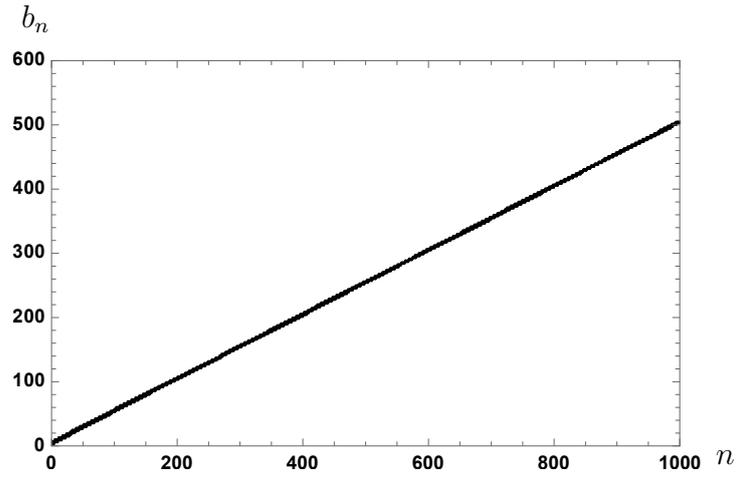
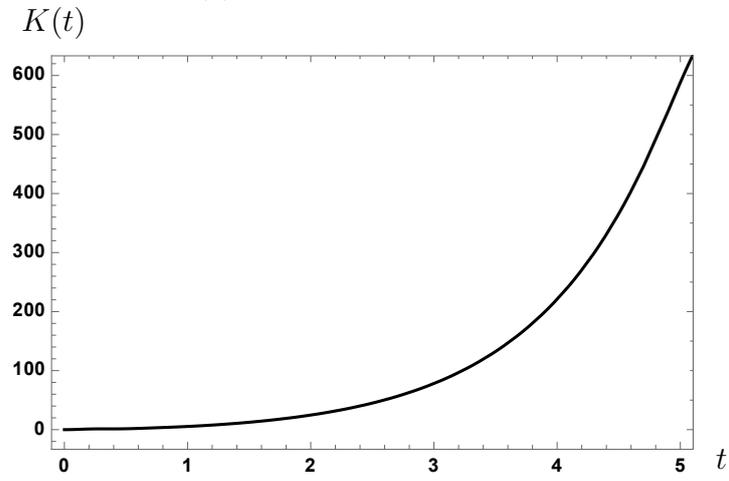
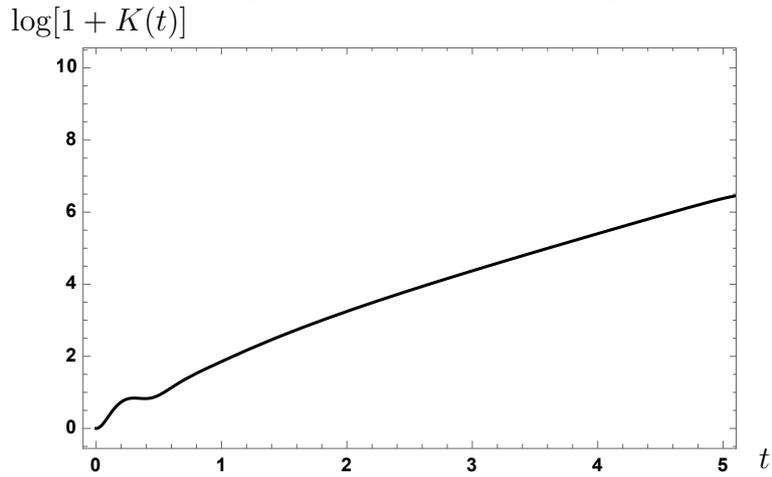

From the holographic computation on the black hole background, the asymptotic behavior of $G_{12}(\omega)$ is determined as (\ref{asbG12BH}). Thus, we consider the following continuous spectrum 
\begin{align}\label{G12N4BH}
G_{12}(\omega)=\frac{1}{N_0}\vert\omega\vert^{2 \nu} e^{  -\frac{\vert\omega\vert \beta}{2}},
\end{align}
and compute the Krylov complexity. 
Figure \ref{fig:KCN4BH} shows the Lanczos coefficient $b_n$ and the Krylov complexity $K(t)$ of the spectrum (\ref{G12N4BH}) for $\nu=10$ and $\beta=2\pi$. To see the growth behavior, we plot $K(t)$ and $\log[1+K(t)]$. One can see the linear growth of $b_n$ with one slope and the linear growth of $\log[1+K(t)]$, which means the exponential growth of $K(t)$, and its exponential growth rate is $K(t)\sim e^{\frac{2\pi}{\beta} t}$ due to $e^{  -\frac{\vert\omega\vert \beta}{2}}$ in (\ref{G12N4BH}).

We comment on the sum for $n$ in numerical computation of $K(t)= \sum_{n=1}^\infty n|\varphi_n(t)|^2$. In our numerical computation for Figure \ref{fig:KCN4BH}, we only calculate the Lanczos coefficient $b_n$ up to $n_{max}=1000$ and evaluate $K(t)\sim\sum_{n=1}^{n_{max}} n|\varphi_n(t)|^2$. Under this approximation, $K(t)$ is bounded as $K(t)\le n_{max}$. To see the growth of $K(t)$ at much later times, we need to choose larger $n_{max}$ in the numerics. On the other hand, we numerically confirmed that $K(t)$ of the discrete spectrum $G_{12}(\omega)$ (\ref{G12N4thermalAdS}) in Figure \ref{fig:KCN4thermalAdS} does not change significantly when we increase $n_{max}$.\\

Let us summarize our results of $\mathcal{N}=4$ $SU(N)$ Super Yang-Mills theory in the large $N$ limit.
Figures \ref{fig:KCN4thermalAdS} and \ref{fig:KCN4BH} show the two distinct behaviors of $K(t)$: oscillation for thermal AdS (\ref{thermalAdSmetric}) and exponential growth for AdS$_5$ black hole (\ref{AdS5BH}). From these two behaviors of $K(t)$, the Krylov complexity can be interpreted as an order parameter of the Hawking-Page transition that is dual to a confinement/deconfinement phase transition in the large $N$ quantum field theory side. From the viewpoint of spectrum, the different behaviors of Krylov complexity come from the difference between discrete spectrum (\ref{G12N4thermalAdS}) and continuous spectrum (\ref{G12N4BH}). In the thermal AdS geometry (\ref{thermalAdSmetric}), there is only one length scale: the AdS scale $R=1$. Therefore, the temperature scale at which the phase transition occurs and the discrete momentum scale are determined by $1/R=1$ only.

\newpage

\clearpage

\subsection{Krylov complexity for $\mathcal{N}=0$ pure Yang-Mills theory}

We consider the pure $SU(N)$ Yang-Mills theory in the large $N$ limit as a theory with a confinement/de-containment phase. Furthermore, unlike the previous subsection, this theory does not have conformal symmetry. In holography, this is realized by $N$ D4-branes wrapping $\mathbb{S}^1$ circle with an anit-periodic boundary condition for fermions, since then fermions acquire the mass and through one-loop, and bosons also acquire the mass \cite{Witten:1998zw}. See \cite{Csaki:1998qr,deMelloKoch:1998vqw,Brower:1999nj,Brower:2000rp} for calculations of glueball mass concerning the following calculations.\\

\noindent{\bf Setting }

\noindent The above configuration, in the large $N$ limit, can be described by the following metric, 
\begin{align}
\label{AdSsoliton}
ds^2=\left(\frac{U}{R}\right)^{3/2}\left(-dt^2 + \sum_{i=1}^3(dx^i)^2+f(U)dx_4^2 \right)+\left(\frac{R}{U}\right)^{3/2}\left(\frac{dU^2}{f(U)}+U^2 d\Omega_4\right),
\end{align}
where
\begin{align}
\label{ffunctionsoliton}
f(U) = 1 - \frac{U^3_{KK}}{U^3}.
\end{align}
This is the metric for D4-brane wrapping on the thermal circle $x_4$. Here $x_4$ must satisfy the 
periodicity 
\begin{align}\label{AdSSolitonbc1}
x_4 = x_4 + \frac{4 \pi}{3} \frac{R^{3/2}}{U_{KK}^{1/2}},
\end{align}
such that at $U= U_{KK}$, there is no conical deficit. In other words, $(U, x_4)$ coordinate represents flat 2-dimensional space at $U=U_{KK}$, like the tip of the cigar geometry. We also impose a periodic boundary condition 
\begin{align}\label{AdSSolitonbc2}
x^i = x^i+L.
\end{align}

Note that the metric eq.~\eqref{AdSsoliton} is AdS ``soliton" solution, obtained from large $N$ D4-brane black hole metric 
\begin{align}
\label{AdSBH}
ds^2=\left(\frac{U}{R}\right)^{3/2}(- f(U)dt^2 +  \sum_{i=1}^3(dx^i)^2 + d x_4^2)+\left(\frac{R}{U}\right)^{3/2}\left(\frac{dU^2}{f(U)}+U^2 d\Omega_4\right),
\end{align}
by the double Wick rotation $t \to -i x_4$, $x_4 \to + i t$, where $R^3 \propto N$. 
Here $f(U)$ is 
\begin{align}
\label{ffunctionBH}
f(U) = 1 - \frac{U^3_{0}}{U^3},
\end{align}
where $U=U_0$ is a horizon
and the black hole temperature is given by 
\begin{align}
\label{AdSBHper}
\frac{1}{T} = \beta = \frac{4 \pi}{3} \frac{R^{3/2}}{U_{0}^{1/2}}.
\end{align}
Even though \eqref{AdSsoliton} and \eqref{AdSBH} are related by Wick rotation, their physical implication is quite different.  \\
\begin{enumerate}
\vspace{-5mm}
\item The metric \eqref{AdSsoliton} represents the cigar-like geometry, where there is no black hole horizon, and $g_{tt}$ is negative for any $U$. In  \eqref{AdSsoliton}, there is no geometry at $U < U_{KK}$, and the bulk IR cut-off $U_{KK}$ is associated with the QCD scale. The temperature can be arbitrary,  {\it i.e.,} in Euclidean time $\tau:=it$, the periodicity of $\tau$ can be any value. Since \eqref{AdSBH} has already a periodicity \eqref{AdSBHper}, then $\tau, x_4$ are periodic in both metrics. When these periodicities coincide, the free energies calculated from these Euclidean metrics are the same. Therefore, the Hawking-Page phase transition temperature is $\beta=\frac{4 \pi}{3} \frac{R^{3/2}}{U_{KK}^{1/2}}$. 

\vspace{-2mm}
\item On the other hand, \eqref{AdSBH} is a black hole solution where there is a horizon at $U=U_0$ and therefore it has a definite temperature determined by $U_0$ as \eqref{AdSBHper}. 
\end{enumerate}
Mathematically both blackening factors $f$ behave the same. But in one case $U=U_{KK}$ is a flat space, and in the other case, $U=U_0$ is a horizon. \\

Let's consider the scalar equation of motion in this bulk geometries. 
The equation of motion is
\begin{align}
(\Box-m^2)\phi =\frac{1}{\sqrt{g} } \partial_\mu \left( \sqrt{g} g^{\mu\nu} \partial_\nu \phi  \right) - m^2 \phi =0.
\end{align}
This bulk scalar is coupled to the D4's $\mbox{Tr}[F_{\mu\nu}F^{\mu\nu}]$, which is a ``glueball" operator. Therefore the spectrum of this bulk scalar corresponds to the glueball spectrum. We will solve this equation both A) on the AdS soliton, and B) on the BH separately. 

Furthermore, we always take the ansatz 
\begin{align}
\phi = \exp\left(-i\omega t+ik_ix_i\right) \tilde{\phi}(U),
\end{align}
namely, the wave function depends on $t$, $x_i$, and $U$ only, independent of $x_4$ and $\Omega_4$,  where the momentum $k_i$ is discrete,
\begin{align}
k_i = \frac{2\pi \ell_i}{L} \,.
\end{align}
We always set $R=1$. To restore the AdS scale, we can shift $U \to U/R$, $U_{KK} \to U_{KK}/R$, $U_0 \to U_0/R$.

\subsubsection{AdS soliton case}

For the metric \eqref{AdSsoliton}, $\sqrt{g} \propto U^4$, the equation of motion is 
\begin{align}
\label{EOMforsoliton}
\frac{1}{U^4} \partial_U \left( U^{4+3/2}f(U)  \partial_U \tilde{\phi}  \right) + U^{-3/2} \left(  \omega^2  - k_i^2\right) \tilde{\phi} - m^2  \tilde{\phi}  = 0. 
\end{align}
From the on-shell condition on the boundary, 
\begin{align}
\omega^2 -k_i^2 = M^2,
\end{align}
where $M$ is the mass of the boundary glueball, then 
\begin{align}\label{EOMAdSSoliton}
\frac{1}{U^4} \partial_U \left( U^{4+3/2}f(U)  \partial_U  \tilde{\phi}(U) \right) + \left( \frac{M^2}{U^{3/2}} - m^2 \right) \tilde{\phi}   = 0. 
\end{align}

\noindent{\bf \underline{Near the boundary, $U \to \infty$}}\\
First we solve the equation of motion in the large $U$ region, where $f(U) \to 1$ and $M^2/U^{3/2} \ll m^2$, and the equation becomes asymptotically as 
\begin{align}
\frac{1}{U^4} \partial_U \left( U^{4+3/2}  \partial_U  \tilde{\phi}(U) \right)  - m^2  \tilde{\phi} \approx  0  \quad \mbox{at large $U$}. 
\end{align}
Asymptotic solutions are given by
\begin{align}
\label{solitoneq1}
\tilde{\phi}(U) \approx  c_1 \frac{ I_9\left(4 m U^{1/4}\right)}{\left( 4 m
   U^{1/4} \right)^9}+c_2 \frac{ K_9\left(4 m U^{1/4}\right)}{\left( 4 m
   U^{1/4} \right)^9} \,, \quad \mbox{at large $U$}, 
\end{align}
where $I$ and $K$ are Bessel functions. At large $U$, 
\begin{align}
\label{solitoneq2}
\lim_{U\to \infty} \frac{ I_9\left(4 m U^{1/4}\right)}{\left( 4 m U^{1/4} \right)^9}  \to \infty \,, \quad  \lim_{U\to \infty} \frac{ K_9\left(4 m U^{1/4}\right)}{\left( 4 m  U^{1/4} \right)^9} \to 0,
\end{align}
therefore we need the boundary condition $c_1 = 0$. \\

\noindent{\bf \underline{Near the tip, $U \to U_{KK}$}}\\
On the other hand, near the tip of the cigar $U=U_{KK}$, we have 
$f(U) \approx \frac{3}{U_{KK}} (U - U_{KK})$, the equation of motion reduces to 
\begin{align}
{3  U_{KK}^{1/2} } \partial_U \left(   (U - U_{KK}) \partial_U \tilde{\phi}  \right) + \left(  U_{KK}^{-3/2} M^2  - m^2 \right) \tilde{\phi}  \approx 0.
\end{align}
Asymptotic solutions are given by
\begin{align}
\tilde{\phi}(U) &\approx  c_{tip1} I_0\left(m_G \sqrt{U - U_{KK}} \right) + c_{tip2} K_0\left( m_G \sqrt{U - U_{KK}} \right)  \,,
\end{align}
where
\begin{align}
& m_G \equiv \frac{2}{\sqrt{3 U_{KK}}} \left( m^2 - \frac{M^2}{U_{KK}^{3/2}}   \right)  \,.
\end{align}
Again at the tip of the cigar, 
\begin{align}
\lim_{U\to U_{KK}} I_0\left(m_G \sqrt{U - U_{KK}} \right)  \to 1 \,, \quad  \lim_{U\to U_{KK}}  K_0\left( m_G \sqrt{U - U_{KK}} \right)  \to \infty
\end{align}
therefore we need the boundary condition $c_{tip2} = 0$ at $U=U_{KK}$. 

In summary, we have two boundary conditions $c_1 = 0$ at the boundary and $c_{tip2} = 0$ at the tip. 
However, this is impossible for a general value of $M^2$. By imposing $c_1=0$ at the boundary and extrapolating the solution to $U=U_{KK}$, we obtain the solution with $c_{tip2} = 0$ only for certain values of $M^2$. Just as in ordinary quantum mechanics, this is the reason why $M^2$ is quantized and we obtain the holographic discrete spectrum for scalar glueball operators.

One possible way to determine the values of $M^2$ is the shooting method; numerically to start with the solution at large $U$ with $c_1=0$ and then extrapolate the solution to $U=U_{KK}$. For generic value of $M^2$, $\tilde{\phi}$ diverges so we fine-tune $M^2$ so that at $U=U_{KK}$, $\tilde{\phi}$ converges. Specifically, in the shooting method, we consider the following boundary condition at large $U_{b}$:
\begin{align}
\tilde{\phi}(U_b) =c_2 \frac{ K_9\left(4 m U_b^{1/4}\right)}{\left( 4 m
   U_b^{1/4} \right)^9},\;\;\; \frac{\text{d} \tilde{\phi}(U)}{\text{d}U}\Big\vert_{U=U_b}=c_2 \frac{\text{d}}{\text{d} U}\left[\frac{ K_9\left(4 m U^{1/4}\right)}{\left( 4 m
   U^{1/4} \right)^9}\right]\Bigg\vert_{U=U_b},
   \end{align}
where we choose $c_2$ so that $\tilde{\phi}(U_b)=1$. By solving the EOM (\ref{EOMAdSSoliton}) numerically with this boundary condition, we can compute $\tilde{\phi}(U_{KK}+\epsilon)$ for a given value of $M$, where $\epsilon$ is a small constant. 

\begin{figure}
         \centering
         \includegraphics[width=0.7\textwidth]{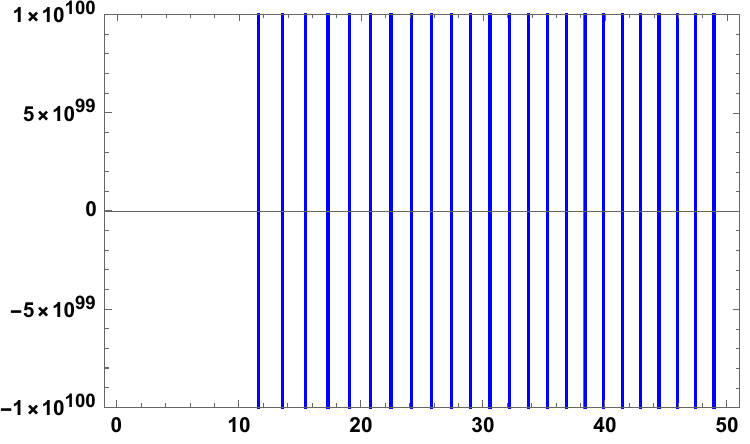}
       \put(5,10){$M$}
    \put(-305,185){$\tilde{\phi}(U_{KK}+\epsilon)$}
             \caption{$M$-dependence of $\tilde{\phi}(U_{KK}+\epsilon)$ for $U_{KK}=1$, $m=10$, $U_b=10^4$, $\epsilon=10^{-4}$.}
        \label{fig:ShootingMethod}
\end{figure}

We plot $\tilde{\phi}(U_{KK}+\epsilon)$ for $U_{KK}=1$, $m=10$, $U_b=10^4$, $\epsilon=10^{-4}$ in Figure \ref{fig:ShootingMethod}, where the horizontal axis is $M$. For generic value of $M$, $\vert\tilde{\phi}(U_{KK}+\epsilon)\vert$ is very large, which means the divergence of $\tilde{\phi}$ at $U=U_{KK}$. However, for some quantized value of $M$, $\vert\tilde{\phi}(U_{KK}+\epsilon)\vert$ is zero, which means that $\tilde{\phi}$ for such a value of $M$ converges at $U=U_{KK}$. 

In the same way as example 3 in Section \ref{sec2}, if $\beta/L\to\infty$, only $k_i =\frac{2\pi \ell_i}{L}=0$ mode is dominant in the spectrum. As $\beta/L$ decreases, the spectrum includes a correction by nonzero $\ell_i$ modes. 
As well as (\ref{G12N4thermalAdS}) for thermal AdS, we first consider the following discrete spectrum for $\beta/L\to\infty$ 
\begin{align}\label{G12N0AdSsoliton}
G_{12}(\omega)=\frac{1}{N_0}\sum_{n=0}e^{-\frac{\vert\omega\vert \beta}{2}}\left[\delta\left(\omega-M_n\right)+\delta\left(\omega+M_n\right)\right],
\end{align}
where $M_n$ is the quantized value determined by the shooting method, and the sequence of $M_n$ is taken as follows
\begin{align}
M_{n_1}<M_{n_2} \;\; \text{if} \;\; n_1<n_2.
\end{align}
For example, $M_1-M_0\sim2$ in Figure \ref{fig:ShootingMethod}.
Next, including the collection of nonzero $k_i = \frac{2\pi \ell_i}{L}$, consider the following discrete spectrum 
\begin{align}\label{G12N0AdSsolitonNonzeroL}
G_{12}(\omega)=&\frac{1}{N_0}\sum_{n=0}\sum_{\ell_i}e^{-\frac{\vert\omega\vert \beta}{2}}\left[\delta\left(\omega-\omega_{n\ell_i}\right)+\delta\left(\omega+\omega_{n\ell_i}\right)\right],\\
\omega_{n\ell_i}:=&\sqrt{M_n^2+\left(\frac{2\pi}{L}\right)^2\left(\ell_1^2+\ell_2^2+\ell_3^2\right)},\label{DispersionRelation}
\end{align}
where $\ell_i$ takes integer values in the sum. If $\beta/L\ll1$, the momentum $k_i = \frac{2\pi \ell_i}{L}$ can be treated as continuous, and the spectrum (\ref{G12N0AdSsolitonNonzeroL}) is close to continuous.

In the AdS soliton geometry, there are two length scales for the periodic boundary conditions (\ref{AdSSolitonbc1}) and (\ref{AdSSolitonbc2}), where $\frac{4 \pi}{3} \frac{R^{3/2}}{U_{KK}^{1/2}}$ is associated to the QCD scale, and $L$ is associated to the discrete momentum $k_i =\frac{2\pi \ell_i}{L}$. The phase transition occurs when $\beta=\frac{4 \pi}{3} \frac{R^{3/2}}{U_{KK}^{1/2}}$. Therefore, for the prescription in Section \ref{sec2}, we should take 
\begin{align}\label{CAdSSoliton}
\beta \sim  \frac{R^{3/2}}{U_{KK}^{1/2}}, \;\;\; \beta/L \gtrsim 1.
\end{align}
In the following numerical computations, we set $R=1$, $U_{KK}=1$, $m=10$, $\beta=2\pi$.

\begin{figure}
\centering
     \begin{subfigure}[b]{0.6\textwidth}
         \centering
         \includegraphics[width=\textwidth]{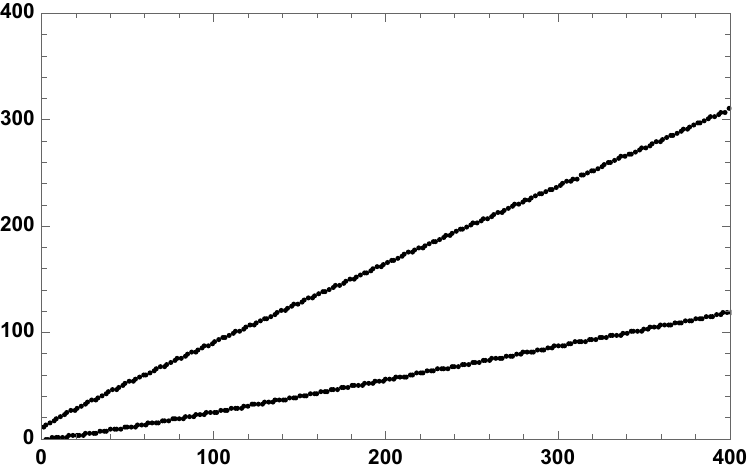}
       \put(5,5){$n$}
    \put(-255,170){$b_n$}
       \caption{$b_n$ of the spectrum (\ref{G12N0AdSsoliton}) for $\beta/L\to\infty$}\label{fig:bnL0}
     \end{subfigure}
          \begin{subfigure}[b]{0.6\textwidth}
         \centering
         \includegraphics[width=\textwidth]{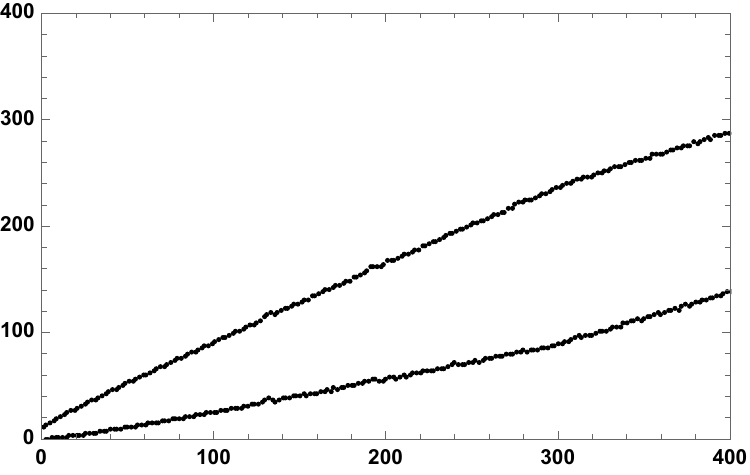}
       \put(5,5){$n$}
    \put(-255,170){$b_n$}
       \caption{$b_n$ of the spectrum (\ref{G12N0AdSsolitonNonzeroL}) for $\beta/L=150$}\label{fig:bnNonzeroL150}
     \end{subfigure}
           \begin{subfigure}[b]{0.6\textwidth}
         \centering
         \includegraphics[width=\textwidth]{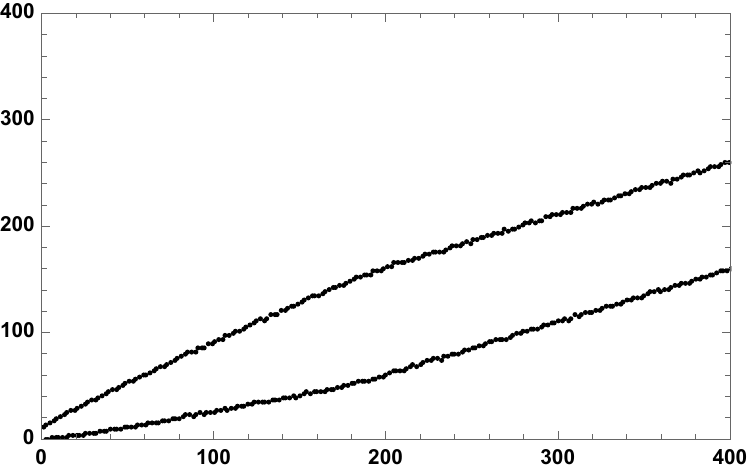}
       \put(5,5){$n$}
    \put(-255,170){$b_n$}
       \caption{$b_n$ of the spectrum (\ref{G12N0AdSsolitonNonzeroL}) for $\beta/L=100$}\label{fig:bnNonzeroL}
     \end{subfigure}
             \caption{Numerical plots of the Lanczos coefficients $b_n$ for $U_{KK}=1$, $m=10$, $\beta=2\pi$.}
        \label{fig:bn}
\end{figure}

 Figure \ref{fig:bnL0} shows the Lanczos coefficient $b_n$ of the spectrum (\ref{G12N0AdSsoliton}) for $\beta/L\to\infty$. One can see the two-slope behavior of $b_n$, which is a characteristic behavior for discrete spectrum like (\ref{twoslope}). Next, Figure \ref{fig:bnNonzeroL150} shows the Lanczos coefficient $b_n$ of the spectrum (\ref{G12N0AdSsolitonNonzeroL}) for $\beta/L=150$. At small $n$, $b_n$ in Figure \ref{fig:bnNonzeroL150} also has the two-slopes behavior. However, from $n\sim300$, the slopes for odd $n$ and even $n$ seem to be the same. Two such identical slopes are observed in a massive free scalar theory in non-compact space \cite{Avdoshkin:2022xuw, Camargo:2022rnt} whose spectrum is continuous. Finally, Figure \ref{fig:bnNonzeroL} shows the Lanczos coefficient $b_n$ of the spectrum (\ref{G12N0AdSsolitonNonzeroL}) for $\beta/L=100$, where the two slopes behavior ends at $n\sim200$.  These figures show that, as the length scale $L$ of compact space increases, the behavior of Lanczos coefficient $b_n$ approaches the behavior for a continuous spectrum at smaller $n$ due to the momentum $k_i = \frac{2\pi \ell_i}{L}$.
 
\begin{figure}
\centering
     \begin{subfigure}[b]{0.6\textwidth}
         \centering
         \includegraphics[width=\textwidth]{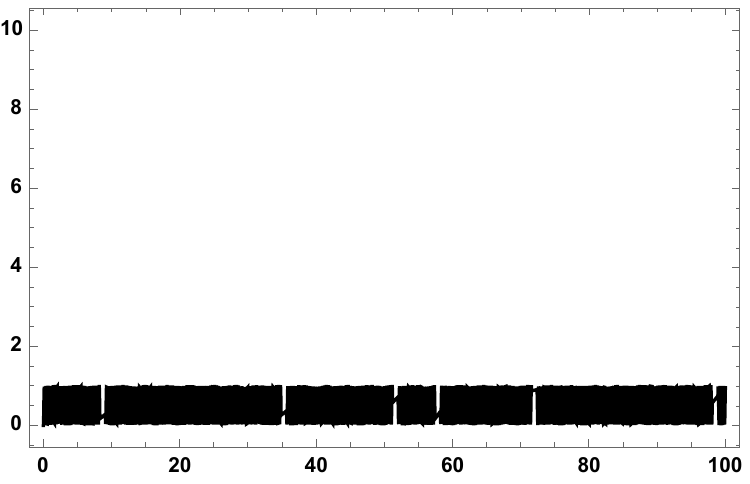}
       \put(5,5){$t$}
    \put(-255,170){$K(t)$}
      \caption{Krylov complexity $K(t)$ of the discrete spectrum $G_{12}(\omega)$ (\ref{G12N0AdSsoliton}) for $\beta/L\to\infty$.}
        \label{fig:KCN0AdSsolitonL0}
     \end{subfigure}
            \begin{subfigure}[b]{0.6\textwidth}
         \centering
         \includegraphics[width=\textwidth]{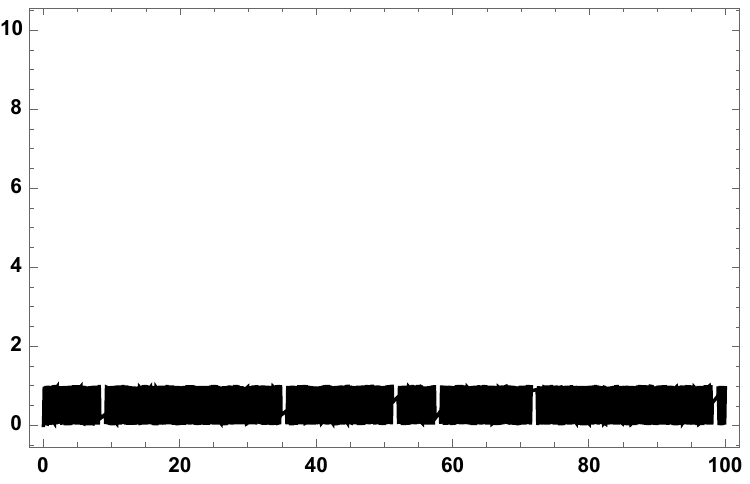}
       \put(5,5){$t$}
    \put(-255,170){$K(t)$}
      \caption{Krylov complexity $K(t)$ of the discrete spectrum $G_{12}(\omega)$ (\ref{G12N0AdSsoliton}) for $\beta/L=150$.}
        \label{fig:KCN0AdSsolitonL150}
     \end{subfigure}
                 \begin{subfigure}[b]{0.6\textwidth}
         \centering
         \includegraphics[width=\textwidth]{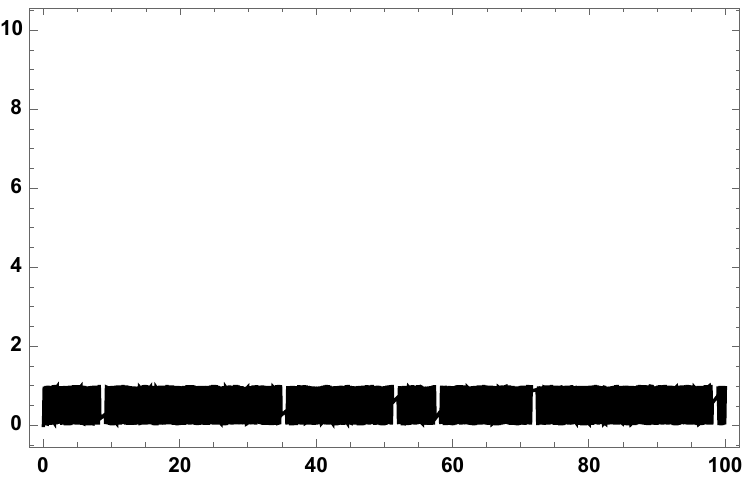}
       \put(5,5){$t$}
    \put(-255,170){$K(t)$}
      \caption{Krylov complexity $K(t)$ of the discrete spectrum $G_{12}(\omega)$ (\ref{G12N0AdSsoliton}) for $\beta/L=100$.}
        \label{fig:KCN0AdSsolitonL100}
     \end{subfigure}
             \caption{Numerical plots of the Krylov complexity $K(t)$ for $U_{KK}=1$, $m=10$, $\beta=2\pi$.}
        \label{fig:KCSmallL}
\end{figure}

Figure \ref{fig:KCN0AdSsolitonL0} shows the Krylov complexity $K(t)$ computed from the discrete spectrum $G_{12}(\omega)$ (\ref{G12N0AdSsoliton}) for $\beta/L\to\infty$, where $K(t)$ oscillates and does not grow. In particular, the maximum value of $K(t)$ is $K(t)\sim1$, which is similar to (\ref{KCHO}). Figures \ref{fig:KCN0AdSsolitonL150} and \ref{fig:KCN0AdSsolitonL100}  shows $K(t)$ computed from the discrete spectrum $G_{12}(\omega)$ (\ref{G12N0AdSsolitonNonzeroL}) for $\beta/L=150,100$. At least in the time region $0\le t \le100$, which is larger than the time scale $t\sim\beta=2\pi$, these three figures are identical. In these figures, the values of dimensionless scales measuring the discreteness of (\ref{G12N0AdSsolitonNonzeroL}) are 
\begin{align}
\beta (M_1-M_0) \sim 4\pi \sim O(10), \;\;\; \beta/L \gtrsim O(100).
\end{align}
Due to these not-small values, $K(t)$ shows the oscillation behavior.

\newpage

\clearpage

\newpage

\clearpage

\begin{figure}
\centering
     \begin{subfigure}[b]{0.6\textwidth}
         \centering
         \includegraphics[width=\textwidth]{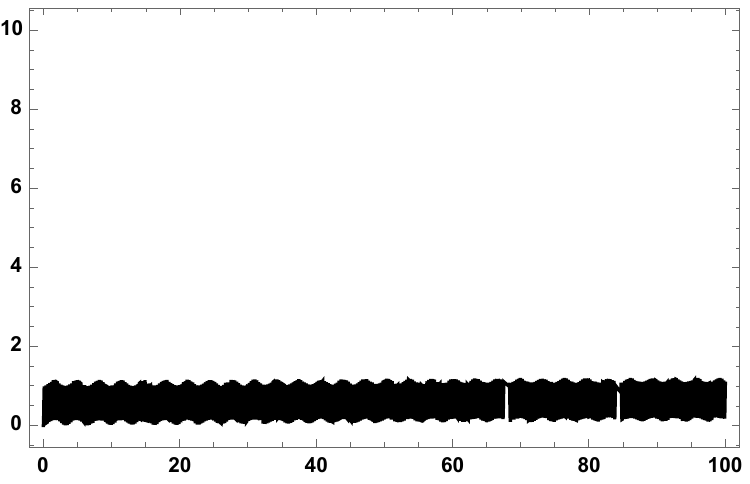}
       \put(5,5){$t$}
    \put(-255,170){$K(t)$}
      \caption{Krylov complexity $K(t)$ of the discrete spectrum $G_{12}(\omega)$ (\ref{G12N0AdSsolitonNonzeroL}) for $\beta/L=7$.}
        \label{fig:KCN0AdSsolitonNonzeroL7}
     \end{subfigure}
          \begin{subfigure}[b]{0.6\textwidth}
         \centering
         \includegraphics[width=\textwidth]{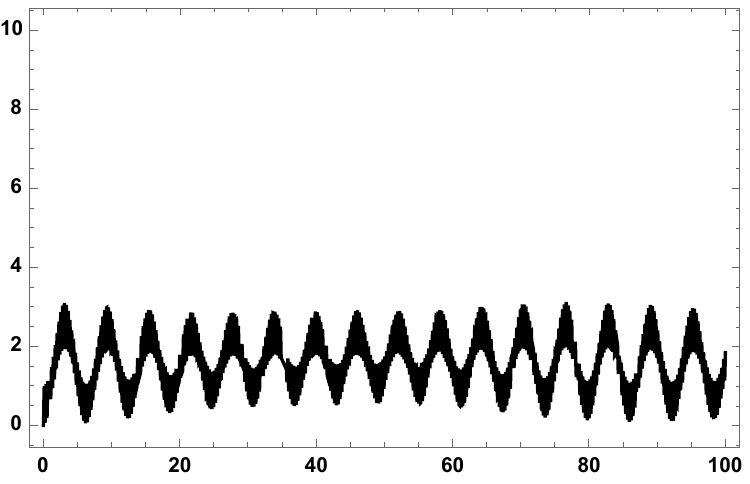}
       \put(5,5){$t$}
    \put(-255,170){$K(t)$}
             \caption{Krylov complexity $K(t)$ of the discrete spectrum $G_{12}(\omega)$ (\ref{G12N0AdSsolitonNonzeroL}) for $\beta/L=5$.}
        \label{fig:KCN0AdSsolitonNonzeroL5}
     \end{subfigure}
             \caption{Numerical plots of the Krylov complexity $K(t)$ for $U_{KK}=1$, $m=10$, $\beta=2\pi$.}
        \label{fig:KCLargeL}
\end{figure}

Three figures of $K(t)$ in Figure \ref{fig:KCSmallL} are identical since $\beta/L$ is too large as $\beta/L \gtrsim O(100)$. If $\beta/L \sim \order{1}$, $K(t)$ would depend on the value of $\beta/L$.
Figure \ref{fig:KCLargeL} shows $K(t)$ computed from the discrete spectrum $G_{12}(\omega)$ (\ref{G12N0AdSsolitonNonzeroL}) for $U_{KK}=1$, $m=10$, $\beta=2\pi$, $\beta/L=7,5$. Compared to $K(t)$ in Figure \ref{fig:KCSmallL}, the maximum values of $K(t)$ in Figure \ref{fig:KCLargeL} are larger due to $\beta/L \sim \order{1}$. The values of $t$ at which the Krylov complexity is maximized change as $\beta/L$ changes since the eigenvalue interval in the discrete spectrum changes. Our numerical results show the validity of prescription II in Section \ref{sec2}. By compactifying the space sufficiently $\beta/L \gtrsim 1$, the exponential growth of $K(t)$ due to continuous momentum does not occur, and thus $K(t)$ oscillates and does not grow if the mass spectrum is discrete.

\subsubsection{BH case}

Similarly for the BH background metric \eqref{AdSBH}, 
with $\sqrt{g} \propto U^4$, the equation of motion is 
\begin{align}
\frac{1}{U^4} \partial_U \left( U^{4+3/2}f(U)  \partial_U \tilde{\phi}  \right) + U^{-3/2} \left( \frac{\omega^2}{f(U)}   - k_i^2\right) \tilde{\phi} - m^2  \tilde{\phi}  = 0. 
\end{align}
Additional warping factor in front of $\omega^2$ is the crucial difference between this black hole metric and \eqref{EOMforsoliton} for AdS soliton. 
Again before we use the tortoise coordinate and make the equation into the form of Schrodinger equation. Then, let us analyze the boundary condition. \\

\noindent{\bf \underline{Near the boundary, $U \to \infty$}}\\
In this case, the analysis is the same as the AdS soliton case. The solution is given by 
\eqref{solitoneq1} and \eqref{solitoneq2}, and we need the boundary condition $c_1=0$.\\

\noindent{\bf \underline{Near the horizon, $U \to U_{0}$}}\\
The crucial difference appears here. Near the horizon, we have  
$f(U) \approx \frac{3}{U_{0}} (U - U_{0})$, and $\omega^2/f(U) \gg k_i^2, m^2$. The equation of motion reduces to 
\begin{align}
&{3  U_{0}^{1/2} } \partial_U \left(   (U - U_{0}) \partial_U \tilde{\phi}  \right) + \left(  U_{0}^{-3/2} \left(\frac{U_0 \, \omega^2}{3 (U - U_0)} - k_i^2 \right)  - m^2 \right) \tilde{\phi}   \\
= \, & {3  U_{0}^{1/2} } \partial_U \left(   (U - U_{0}) \partial_U \tilde{\phi}  \right) + \frac{ \omega^2}{3U_{0}^{1/2}  (U - U_0)} \tilde{\phi}   \,  \approx 0  \quad \mbox{at $U \to U_0$},
\end{align}
which allows 
\begin{align}
\tilde{\phi}(U) &\approx   c_{bhc} \cos \left(\frac{\omega \log (U-U_0)}{3
   \sqrt{U_0}}\right)+c_{bhs} \sin \left(\frac{\omega \log (U-U_0)}{3
   \sqrt{U_0}}\right) \\
   & = c_{bh1} \exp \left( {i \, \frac{\omega}{3 \sqrt{U_0}} \log (U - U_0)) }  \right) +  c_{bh2} \exp \left( -{i \, \frac{\omega}{3 \sqrt{U_0}} \log (U - U_0)) }  \right).
   \label{nearhorizon1}
\end{align}
Even though both solutions oscillate very heavily near the horizon, there is no divergence. So we can set both $c_{bh1}$ and $c_{bh2}$ to be nonzero consistently. Again the crucial difference is $f(U)$ in front of $\omega$, which is due to $g_{tt}$. 

Therefore by numerically starting with the solution at large $U$ with $c_1 = 0$, we can extrapolate the solution to the horizon $U = U_0$. For the generic value of $\omega$ without any fine-tuning, $\tilde{\phi}$ converges even at the horizon. This is the reason why $\omega$ takes continuous value for the black hole case. 
Intuitively it is clear that there are localized waves near the horizon whose energy is extremely low due to the warping factor of $g_{tt} \to 0$.  This is the reason why the spectrum becomes continuous even for the compact space in the presence of a black hole horizon. 

To solve the equation, 
let us use the tortoise coordinate;
\begin{align}
\label{tortoise}
z = \int_U^\infty \left( \frac{R}{U} \right)^{3/2} \frac{dU}{f(U)}.
\end{align}
Then, $z$-derivative becomes $-(U/R)^{3/2}f(U)\partial_U=\partial_z$, and we decompose $\phi$ into
\begin{align}
\phi=\exp\left(-i\omega t+ik_ix_i\right)U^{\alpha}\psi(U)
\end{align}
By tuning $\alpha = -2$, one can make the equation in the form of a Schrodinger equation. 
Then the equation of motion becomes 
\begin{align}
\label{Schrodinger1}
& -\partial_z^2 \psi(U)     + V(U) \psi(U)= \omega^2\psi(U) \\
{\rm where}&\quad V(U)=f(U) \left[ 5Uf(U)+2U^2f'(U)+m^2U^{3/2} +  \vec{k}^2 \right] \,,
\label{Schrodinger2}
\end{align}
where $\vec{k}=\frac{2\pi}{L}(n_x,n_y,n_z)$ is quantized momentum.
At $U\to \infty$, the mass term potential dominates
\begin{align}
\label{massboxp4}
\lim_{U \to \infty}V(U) \approx  m^2 U^{3/2}.
\end{align}
This is essential because AdS is a gravitational box.
Therefore we choose the boundary condition so that $\phi \propto \psi(U)/U^2$  at $U\to\infty$  converges as we have seen previously. 

At the horizon $U=U_0$,  
$V(U) \to 0$ therefore we can normalize it to be real as \eqref{bconaround0SAdS} 
\begin{align}
\label{bconaround0}
\psi(U) = e^{i \omega z - i \delta} +  e^{-i \omega z + i \delta} \,.
\end{align}
From \eqref{tortoise}, as we increase $U$, $z$ decreases. Therefore, 
\begin{enumerate}
\item $ e^{- i \omega z - i \delta} $, which corresponds to $ \exp \left( {i \, \frac{\omega}{3 \sqrt{U_0}} \log (U - U_0)) }  \right)$, is an out-going mode.
\item $ e^{+ i \omega z - i \delta} $, which corresponds to $ \exp \left(- {i \, \frac{\omega}{3 \sqrt{U_0}} \log (U - U_0)) }  \right)$, is an in-going mode. 
\end{enumerate}
As \cite{Festuccia:2005pi}, we consider the Wightman Green function and thus kept both ingoing and outgoing modes.

We can now solve this by using the WKB approximation, where $m\to \infty$. We will read the large $\omega$ behavior of the spectral function for the Krylov complexity. \\

\noindent{\bf \underline{WKB approximation}}\\
We turn our attention to the following scaling limit 
\begin{align}
\label{largenu}
\omega =\nu u \,,\quad m  = \nu \,, \quad  \vec{k} = \nu \vec{l} \,, \quad \nu \gg 1.
\end{align}
By setting 
\begin{align}
\psi(U)=e^{\nu S}\,,  
\end{align}
from \eqref{Schrodinger1}, we have 
\begin{align}
-\nu (\partial_z^2 S) -\nu^2(\partial_z S)^2+\nu^2 V_0  =\nu^2 u^2 + \order{\nu^0},
\end{align}
where $V_0$ is 
\begin{align}
V_0 &= f(U) \left(U^{3/2}  + \vec{l}^2 \right)  \,.
\end{align}
This  $V_0$ is obtained from $V$ in eq.~\eqref{Schrodinger2} as  
\begin{align}
V &=f(U) \left[ 5Uf(U)+2U^2f'(U)+m^2U^{3/2} +  \vec{k}^2 \right] \\
&= \nu^2 V_0 + \order{ \nu^0} \,.
\end{align}

Setting
\begin{align}
 S=S^{(0)}+\frac{1}{\nu}S^{(1)}+\cdots\,, 
\end{align}
for the WKB approximation, 
leading and subleading order $S^{(0)}$, $S^{(1)}$ satisfies
\begin{align}
-\nu (\partial_z^2 S^{(0)}+\frac{1}{\nu} \partial_z^2 S^{(1)})-\nu^2(\partial_z S^{(0)}+\frac{1}{\nu} \partial_z S^{(1)})^2+\nu^2 V_0=\nu^2 u^2 + \order{\nu^0}.
\end{align}
Therefore in the leading order at $\order{\nu^2}$,
\begin{align}
-&(\partial_z S^{(0)})^2+V_0= u^2 \ \ \ \to\ \ \ \partial_z S^{(0)}=+\sqrt{V_0-u^2} \\
& \to\ \ \ S^{(0)}  = \int dz \sqrt{V_0 - u^2} = -\int_{U_c}^{U} dU' \kappa(U') \,, \\
&\mbox{where} \quad \kappa=\left( \frac{R}{U} \right)^{3/2} \frac{1}{f(U)} \sqrt{V_0-u^2} \,.
\end{align}
Here we used \eqref{tortoise} and 
\begin{align}
\label{turningpt}
V_0(U_c) = u^2 \,,
\end{align} 
at the turning point $U = U_c$. 
At the subleading order $\order{\nu}$,
\begin{align}
 -\partial_z^2 S^{(0)}-2 \partial_z S^{(0)}\partial_z S^{(1)}=0\ \ \ \to\ \ \ S^{(1)}= \log \frac{1}{\left({ V_0 - u^2} \right)^{1/4}}.
\end{align}
Therefore under the WKB approximation, the approximate solution becomes
\begin{align}
\psi^{(wkb)}(U)= \frac{1}{\left({ V_0 - u^2} \right)^{1/4}} e^{\nu \mathcal{Z}} \left(1+\order{\nu^{-1}} \right),\ \ \ \mathcal{Z}=-\int_{U_c}^{U} dU' \kappa(U') \,.
\end{align}
Near the horizon $V_0 \to 0$ due to $f(U) \to 0$, the normalization condition eq.~\eqref{bconaround0} determines the relative normalization factor as follows
\begin{align}
\psi^{(wkb)}(U)=\frac{1}{\sqrt{u}}\psi(U).
\end{align}

Now, the bulk correlator is
\begin{align}
\mathcal{G}_+ \sim \frac{1}{\omega}(UU')^{-2}\psi(U)\psi(U').
\end{align}
From \eqref{solitoneq2}, the normalizable mode approaches asymptotically
\begin{align}
 \lim_{U\to \infty} \frac{ K_9\left(4 m U^{1/4}\right)}{\left( 4 m  U^{1/4} \right)^9} \to \sqrt{\frac{\pi}{2(4m U^{1/4})}}\frac{e^{-4m U^{1/4}}}{(4mU^{1/4})^9} \,\to 0 \,,\ \ \ \left( \lim_{z\to \infty}K(z)= \sqrt{\frac{\pi}{2z}}e^{-z}\right).
\end{align}
From this, the relationship between boundary correlation function $G_+$ and bulk correlator $\mathcal{G}_+$ is
\begin{align}
G_+ &\sim \lim_{U,U'\to\infty}(e^{4m U^{1/4}} {(4mU^{1/4})^{19/2}})(e^{4m U'^{1/4}} {(4mU'^{1/4})^{19/2}}) \mathcal{G}_+(U,U')\\
& \sim \lim_{U,U'\to\infty}(e^{4m U^{1/4}} {(4mU^{1/4})^{19/2}})(e^{4m U'^{1/4}} {(4mU'^{1/4})^{19/2}}) (UU')^{-2}\notag \\
&\ \ \ \ \ \ \ \ \ \ \ \ \ \ \ \ \ \ \ \ \ \ \ \ \ \ \ \ \ \ \ \ \ \ \  \times \frac{1}{\nu}\frac{1}{ \left(V_0 - u^2 \right)^{1/4} } e^{\nu \mathcal{Z}}\frac{1}{ \left(V_0 - u^2 \right)^{1/4} } e^{\nu \mathcal{Z}}.
\end{align}
Now $\left(V_0 - u^2 \right)^{1/4} \to U^{3/8}$ at large $U$,  then
\begin{align}
G_+&\sim \lim_{U,U'\to\infty}(e^{4m U^{1/4}} {(4mU^{1/4})^{19/2}})(e^{4m U'^{1/4}} {(4mU'^{1/4})^{19/2}}) (UU')^{-2}\frac{1}{\nu}\frac{1}{U^{3/8}} e^{\nu \mathcal{Z}}\frac{1}{U'^{3/8}} e^{\nu \mathcal{Z}}\\
&\sim \lim_{U,U'\to\infty}(4m)^{19}\frac{1}{\nu} e^{ {2\nu (4U^{1/4}+\mathcal{Z})}}.
\end{align}
The argument of the exponent is 
\begin{align}
4U^{1/4}+\mathcal{Z}&=4U^{1/4}-\int_{U_c}^{U} dU' \left( \frac{1}{U'} \right)^{3/2} \frac{1}{f(U')} \sqrt{V_0-u^2}.
\end{align}

To evaluate the Lanczos coefficient at large $n$, we need to evaluate the Green function at large $\omega$. However, at large $\omega$, the turning point $U_c$ becomes large from \eqref{turningpt}.  Therefore in such a large $U_c$, $f(U) \approx 1$, $V_0 \approx U^{3/2}$, and $U_c \approx u^{4/3}$. Then we have 
\begin{align}
\lim_{U \to \infty} \left[ 4U^{1/4}+\mathcal{Z} \right]
&\simeq \lim_{U \to \infty} \left[ 4U^{1/4}- \int_{U_c=u^{4/3}}^{U} dU'  \frac{\sqrt{U'^{3/2}-u^2}}{U'^{3/2}}  \right]\\
&=  \lim_{U \to \infty} \left[ 4U^{1/4} -u^{1/3} \int_1^{u^{-4/3} U} dx \frac{\sqrt{x^{3/2}-1}}{x^{3/2}}  \right] \\
&=  \lim_{U \to \infty} \left[ 
4U^{1/4} -u^{1/3} \left(  4 \left(u^{-4/3} U \right)^{1/4} - \frac{9 \sqrt{3 \pi } \Gamma \left(\frac{5}{3}\right)}{\Gamma   \left(\frac{1}{6}\right)} + \order{U^{-1/4}} \right)  \right] \\
&= \frac{9 \sqrt{3 \pi } \Gamma \left(\frac{5}{3}\right)}{\Gamma   \left(\frac{1}{6}\right)}  u^{1/3} =C(\nu)  \, \omega^{1/3},
\end{align}
where 
\begin{align}
C(\nu)  \equiv \frac{9 \sqrt{3 \pi } \Gamma \left(\frac{5}{3}\right)}{\Gamma   \left(\frac{1}{6}\right)}  \,  \frac{1}{\nu^{1/3}} \,. 
\end{align}
Thus the $G_+(\omega)$ in the large $\omega$ limit becomes 
\begin{align}
\label{Gplusourcase}
G_+(\omega)\propto  e^{  C(\nu)  \omega^{1/3}}  \quad \mbox{at large $\omega$}.
\end{align}
where $\nu = m$. 
Therefore the spectrum grows exponentially in $\omega$. 

As is studied in \cite{Festuccia:2005pi}, for correlator with UV regularization as 
\begin{align}
G_{12}(t) = \Tr \left[ e^{-\beta H} {\cal{O}}(t - i\beta/2)  {\cal{O}}(0) \right].
\end{align}
In frequency space, they are related as 
\begin{align}
G_{12}(\omega) = e^{-\frac{\omega \beta}{2}} G_+(\omega).
\end{align}
This can be understood simply by changing the contour of the $t$ integration of the Fourier transformation.  Then 
\begin{align}
\label{G12ourcase}
G_{12}(\omega)\propto  e^{  C(\nu)  \omega^{1/3} -\frac{\omega \beta}{2}}  \to 0 \quad \mbox{at large $\omega$} \,,
\end{align}
goes to zero at large $\omega$, {\it i.e.,} UV regulated. 

A few comments are in order;
\begin{enumerate}
\item Eq.~\eqref{G12ourcase} is our final result for the ${\cal{N}}=0$ holographic QCD, {\it i.e.},  glueball Green function at large $\omega$ in the holographic bulk D4 on $S^1$ geometry at the deconfinement phase. 
\item For D3-brane AdS$_5$ case, eq.~(4.23) of \cite{Festuccia:2005pi} showed that 
\begin{align}
G_{12}(\omega)\propto \omega^{2 \nu} e^{  -\frac{\omega \beta}{2}}  \to 0 \quad \mbox{at large $\omega$} \,.
\end{align}
\item However for D4 on $S^1$ case, our result is eq.~\eqref{G12ourcase}. The difference is that there is additional $\omega^{1/3}$ power in the argument of the exponential. Although this effect is subleading,  
eq.~\eqref{Gplusourcase} shows that there is a scale from the exponential. On the other hand, eq.~(4.15) in \cite{Festuccia:2005pi} shows there is no scale in AdS$_5$ case. 
\item 
Given this, we expect that if we consider the theory dual to large $N$ QCD, generically it behave as  
\begin{align}
\label{G12generalcase}
G_{12}(\omega)\propto \omega^b e^{  A (\omega/\nu)^{a} -\frac{\omega \beta}{2}}  \to 0 \quad \mbox{at large $\omega$} \,,
\end{align}
where $a$ and $b$ are some constant with $a < 1$. Eq.~\eqref{Gplusourcase}  is a special case of $a=1/3$, $b=0$. \\

\end{enumerate}
Given the generic structure \eqref{G12generalcase} in black hole case for holographic QCD, one can read off the Lanczos coefficients and see how the Krylov complexity should grow as a function of $t$.

\begin{figure}
\centering
  \begin{subfigure}[b]{0.6\textwidth}
         \centering
         \includegraphics[width=\textwidth]{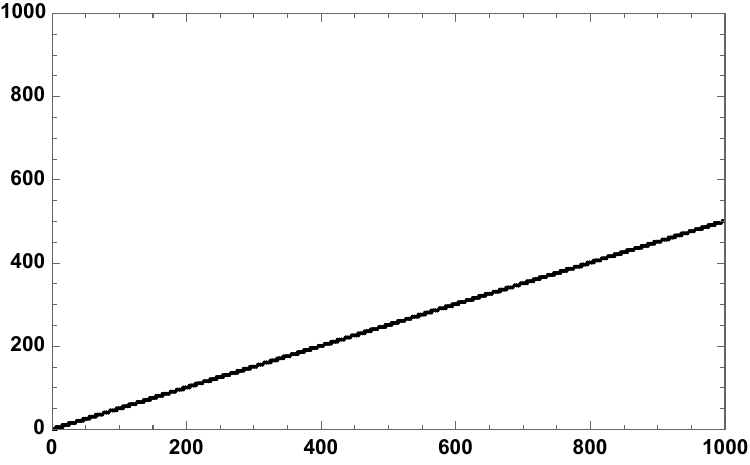}
       \put(5,5){$n$}
    \put(-255,170){$b_n$}
       \caption{Lanczos coefficient $b_n$}\label{fig:bnN0BHb0a13}
     \end{subfigure}
           \begin{subfigure}[b]{0.6\textwidth}
         \centering
         \includegraphics[width=\textwidth]{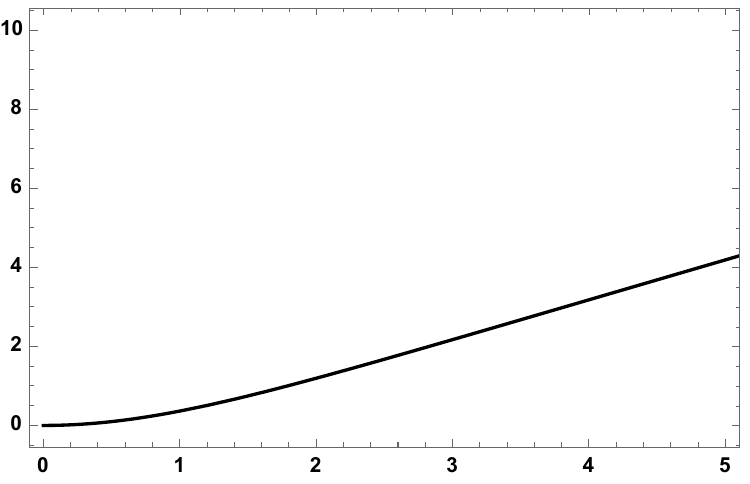}
       \put(5,5){$t$}
    \put(-275,170){$\log[1+K(t)]$}
       \caption{Log plot $\log[1+K(t)]$ of the Krylov complexity}\label{fig:KCN0BHlogb0a13}
     \end{subfigure}
             \caption{Numerical plots of the Lanczos coefficient $b_n$ and the Krylov complexity $K(t)$ of the continuous spectrum $G_{12}(\omega)$ (\ref{G12N0BH}) for $\nu=10$, $\beta=2\pi$, $a=1/3$, $b=0$, $A=\frac{9 \sqrt{3 \pi } \Gamma \left(\frac{5}{3}\right)}{\Gamma   \left(\frac{1}{6}\right)}$, which is for the holographic ${\cal{N}}=0$ pure YM case (\ref{G12ourcase}).
             }
        \label{fig:N0BHb0a13}
\end{figure}

Based on the asymptotic behavior of $G_{12}(\omega)$ (\ref{G12generalcase}), we compute the Krylov complexity of the following continuous spectrum
\begin{align}\label{G12N0BH}
G_{12}(\omega)=\frac{1}{N_0}\vert\omega\vert^{b} e^{ A (\vert\omega\vert/\nu)^{a} -\frac{\vert\omega\vert \beta}{2}},
\end{align}
for the holographic ${\cal{N}}=0$ pure YM case (\ref{G12ourcase}), {\it i.e.}, $a=1/3$, $b=0$, $A=\frac{9 \sqrt{3 \pi } \Gamma \left(\frac{5}{3}\right)}{\Gamma   \left(\frac{1}{6}\right)}\sim4.5$. 
Figure \ref{fig:N0BHb0a13} shows the Lanczos coefficient $b_n$ and the Krylov complexity $K(t)$ of the spectrum (\ref{G12N0BH}) for the holographic ${\cal{N}}=0$ pure YM case with $\nu=10$ and $\beta=2\pi$. The growth behaviors of $b_n$ and $K(t)$ are similar to the behaviors in Figure \ref{fig:KCN4BH}. This is because we set $a=1/3$, and in that case, $A (\vert\omega\vert/\nu)^{a}$ at large $\vert\omega\vert$ in (\ref{G12generalcase}) is smaller than $\frac{\vert\omega\vert \beta}{2}$. Thus, the large $n$ behavior of $b_n$ and the late time behavior of $K(t)$ are mainly determined  from $e^{  -\frac{\vert\omega\vert \beta}{2}}$. 

Just as $\mathcal{N}=4$ case, we implicitly assume the dimension of the operator $\mathcal{O}$ for the Krylov complexity is large enough such that WKB approximation is valid. However, the continuum spectrum from the Green function is determined by the boundary conditions on the horizon thus we expect that the resultant Krylov complexity are not much dependent on the WKB approximation. 
\\

Let us summarize our results of $\mathcal{N}=0$ Yang-Mills theory in the large $N$ limit, where we can arbitrarily choose a length scale $L$ of the compact space. For our prescription in Section \ref{sec2}, we considered AdS soliton geometry (\ref{AdSsoliton}) that satisfies (\ref{CAdSSoliton}).
The Krylov complexity of discrete spectrum (\ref{G12N0AdSsolitonNonzeroL}) for such a geometry oscillates and does not grow. 
In contrast, the Krylov complexity of continuous spectrum (\ref{G12N0BH}) for AdS black hole geometry (\ref{AdSBH}) shows the exponential growth behavior. Together with our results of $\mathcal{N}=4$, the Krylov complexity can be an order parameter of the Hawking-Page transition that is dual to a confinement/deconfinement phase transition at large $N$ by compactifying the space appropriately to avoid continuous momentum.

\section{Summary}

In this paper, we proposed that the Krylov complexity of operators such as $\mathcal{O}=\Tr[F_{\mu\nu}F^{\mu\nu}]$ can be an order parameter of confinement/deconfinement transitions in large $N$ quantum field theories. It has been conjectured that the Krylov complexity grows exponentially in the thermodynamic limit of chaotic systems. However, it is not true that a system is chaotic if the Krylov complexity grows exponentially. An example is free scalar quantum field theories on non-compact space, where continuous momentum in the non-compact space direction is the reason why exponential growth occurs. This implies that the exponential growth of Krylov complexity is sensitive to the continuity of the spectrum. To avoid this exponential growth in free field theory, we can compactify the space on which the field theory is located as $L\lesssim\Lambda_{QCD}^{-1}$ for discrete momentum. Furthermore, if we consider large $N$ field theories, the systems can exhibit confinement/deconfinement transitions. We propose that the exponential growth of Krylov complexity in large $N$ field theories can detect such phase transitions.

Our proposal is inspired by previous studies \cite{Avdoshkin:2022xuw, Kundu:2023hbk, Iizuka:2023pov, Iizuka:2023fba} in which the Krylov complexity behaves in various ways depending on the continuity or discreteness in the spectrum of two-point functions. For a concrete proposal, we formulated conditions of the spectrum that the Krylov complexity grows exponentially: (A) the spectrum is continuous rather than discrete, and (B) the high-energy tail of the spectrum decays exponentially with no upper bound. 

In compact space, the continuity of the spectrum depends on various dimensionful scales such as mass spectrum, temperature, and length scales of the compact space. To use the exponential growth of Krylov complexity as a measure of confinement/deconfinement transitions due to the change of mass spectrum, the spectrum must be prevented from becoming continuous by momentum.

To clearly distinguish the continuity of spectrum due to the KK momentum and mass spectrum, we considered a model of infinitely many free scalars with various masses in compact space. By compactifying the space small enough, one can ignore nonzero discrete momentum, and the continuity of the spectrum is determined by the mass spectrum. Then, the exponential growth of Krylov complexity can be a measure of the continuity of the mass spectrum.

 From the above model, we explicitly proposed the prescription to use the exponential growth of Krylov complexity as a measure of confinement/deconfinement transitions in large $N$ field theories. First, take the temperature near the phase transition scale such as the QCD scale. Next, compactify the space small enough to discretize momentum. Then, the continuity of the spectrum is determined by the continuity of the mass spectrum that changes drastically under the phase transition, and the exponential growth of Krylov complexity can be a measure to detect such change in the spectrum.

As further evidence of our proposal, we studied the Krylov complexity of $\mathcal{N}=4, 0$ $SU(N)$ Yang-Mills theories in the large $N$ limit via holography. First, we analyzed the spectrum of a scalar operator by using a holographic method in \cite{Festuccia:2005pi}. If the background geometry is a black hole geometry, the spectrum is continuous, where we use the WKB approximation to compute the mass spectrum by assuming large scaling dimension. Otherwise, the spectrum is discrete. Then, we calculated the Krylov complexity from the obtained spectrum and confirmed that the behavior of Krylov complexity changes whether the spectrum is continuous or discrete. Therefore, the Krylov complexity can detect the Hawking-Page transitions that are dual to confinement/deconfinement transitions in the large $N$ Yang-Mills theories.

We calculated the Krylov complexity from the spectrum of a bulk scalar that corresponds to the glueball correlator in this paper. But pure Yang-Mills theories contain many other operators. It is a straightforward future work to generalize our computations to the spectrum of bulk gauge or bulk tensor fields and the Krylov complexity of their dual fields.

In principle, the Krylov complexity can be computed numerically if the spectrum of a two-point function is given. It is interesting to evaluate the Krylov complexity from the spectrum in realistic field theories such as QCD. If we do not take the large $N$ limit, a smooth crossover may occur, and the Krylov complexity may also transit smoothly.

In lattice gauge theories, the Yang-Mills action can be approximated by the Wilson action as a sum of plaquette Wilson loops, and a correlation function of two parallel plaquette Wilson loops is used to study a mass gap. Since the Krylov complexity in lattice systems can be defined for $k$-local operators acting on $k$ lattice points, it may be possible to formulate the Krylov complexity for a plaquette Wilson loop operator as a lattice version for $\mathcal{O}=\Tr[F_{\mu\nu}F^{\mu\nu}]$. In the same manner, it may be also possible to formulate the Krylov complexity in quantum field theories for an integral of a local operator on a finite space domain.

\acknowledgments
The works of TA and NI were supported in part by JSPS KAKENHI Grant Number 21J20906(TA), 18K03619(NI). The work of NI is also supported by MEXT KAKENHI Grant-in-Aid for Transformative Research Areas A “Extreme Universe” No. 21H05184. M.N. was supported by the Basic Science Research Program through the National Research Foundation of Korea (NRF) funded by the Ministry of Education (RS-2023-00245035).

\appendix
\section{Spectrum from two-point function}\label{App0}
In this section, we will give a note on the spectrum appearing in the calculation of the Krylov complexity. In general, the Krylov complexity is related to the spectrum $\rho(\omega) = \Re G(\omega)$ determined from a two-point function $G(t)$. Note that this generally does not coincide with the Hamiltonian energy eigenvalues, i.e. the full energy spectrum of the system. Taking the harmonic oscillator 
\begin{align}
H = \frac{p^2}{2m} + \frac{1}{2}m\omega_0^2 x^2,
\end{align}
(full spectrum $H=\hbar \omega_0 (n+\frac{1}{2})$) as an example, we will now calculate the Wightman inner product two-point function and its spectrum given by the Fourier transformation. Starting from
\begin{align}\label{GtHarmonicOscillator}
G(t) &= \frac{1}{Z(\beta)}\Tr \left( e^{-\frac{\beta H}{2}} x(t) e^{-\frac{\beta H}{2}} x(0) \right)\\
&= \frac{1}{Z(\beta)}\Tr \left( e^{-\frac{\beta H}{2}} e^{iHt}x(0)e^{-iHt} e^{-\frac{\beta H}{2}} x(0) \right),
\end{align}
where $Z(\beta)$ is the partition function. Let us write down by using full energy eigenstate $\ket{n}$, 
\begin{align}
G(t) &= \frac{1}{Z(\beta)}\sum_{n,n'} \braket{n| e^{-\frac{\beta H}{2}} e^{iHt}x(0)|n'}\braket{n'| e^{-iHt} e^{-\frac{\beta H}{2}} x(0) |n}\notag \\
&= \frac{1}{Z(\beta)}\sum_{n,n'} e^{-\frac{\beta}{2}(E_n+E_{n'})} e^{i(E_n-E_{n'}) t} \braket{n| x(0)|n'}\braket{n'| x(0) |n}.\notag
\end{align}
From knowledge of harmonic oscillators, the matrix components can be calculated as follows
\begin{align}
 \braket{n'| x(0)|n} &= \sqrt{\frac{\hbar}{2m\omega_0}}\left( \delta_{n',n+1} \sqrt{n+1} + \delta_{n',n-1}\sqrt{n} \right).
\end{align}
From these,
\begin{align}
G(t) &= \frac{1}{Z(\beta)}\frac{\hbar}{2m\omega_0} \sum_{n,n'} e^{-\frac{\beta}{2}(E_n+E_{n'})} e^{i(E_n-E_{n'}) t} (\delta_{n,n'+1} \sqrt{n'+1} \delta_{n',n-1}\sqrt{n} + \delta_{n,n'-1}\sqrt{n'} \delta_{n',n+1} \sqrt{n+1}  )\notag \\
&= \frac{1}{Z(\beta)}\frac{\hbar}{2m\omega_0} \left(\sum_{n=1} n e^{-\frac{\beta}{2}(E_n+E_{n-1})} e^{i(E_n-E_{n-1}) t}  + \sum_{n=0}(n+1) e^{-\frac{\beta}{2}(E_n+E_{n+1})} e^{i(E_n-E_{n+1}) t} \right)\notag\\
&= \frac{1}{Z(\beta)}\frac{\hbar}{2m\omega_0} \sum_{n=1} n e^{-\hbar \omega_0 \beta n} \left(e^{i\hbar \omega_0 t}+e^{-i\hbar \omega_0 t}\right)= \frac{\hbar \cos [\hbar\omega_0 t]}{2m\omega_0 \sinh[\hbar\omega_0\beta/2]} ,  
\end{align}
where we used $E_n=\hbar \omega_0 (n+\frac{1}{2})$ and $Z(\beta) = \Tr (e^{-\beta H}) = \sum_n e^{-\beta \hbar \omega_0 (n+\frac{1}{2})} = \frac{1}{2\sinh[\beta \hbar \omega_0/2]}$. Fourier transformation of this function is
\begin{align}
G(\omega) &= \int dt e^{i\omega t} G(t) = \frac{1}{Z(\beta)}\int dt \frac{\hbar}{2m\omega_0} \sum_{n=1} n e^{-\hbar \omega_0 \beta n} \left(e^{i(\omega + \hbar \omega_0 ) t}+e^{i(\omega - \hbar \omega_0 ) t}\right)\notag \\
&= \frac{\pi\hbar}{2m\omega_0 \sinh[\hbar\omega_0\beta/2]} \left( \delta(\omega + \hbar \omega_0 ) + \delta(\omega - \hbar \omega_0 )\right) .
\end{align}
Since the sum over $n$ is simply a constant, this spectrum is simply a delta-functional distribution with $\omega = \pm \hbar \omega_0$. This is related to the fact that the matrix components of $x$ can only represent transitions between neighboring energy eigenstates. Therefore, the spectrum calculated from the two-point function of $x$, which is different from the full energy spectrum $E_n=\hbar \omega_0(n+\frac{1}{2})$, is also in such a way that only transitions between neighboring energy states are allowed. The Krylov complexity uses information on the spectrum determined from the two-point function. 

\section{Basis of Krylov complexity}\label{AppA}
In this subsection, we will briefly review the definition of Krylov complexity and a related important quantity, Lanczos coefficient. Let us consider a local operator $\hat{\mathcal{O}}$ and its time evolution
\begin{align}
\hat{\mathcal{O}}(t)&=e^{iHt} \hat{\mathcal{O}}e^{-iHt}=\hat{\mathcal{O}}+it[H,\hat{\mathcal{O}}]+\frac{(it)^2}{2!}[H,[H,\hat{\mathcal{O}}]]+\cdots \\
&= \sum_{n=0}^{\infty}\frac{(it)^n}{n!} \mathcal{L}^n \hat{\mathcal{O}} = e^{i\mathcal{L}t}\hat{\mathcal{O}}\, ,\\
\mathcal{L}^n \hat{\mathcal{O}}&\equiv \underbrace{[H,[H,[H, \cdots,[H,\hat{\mathcal{O}}]]]]}_{n \, \, commutators}.
\end{align}
The Krylov complexity measures the spread of $\hat{\mathcal{O}}(t)$ in Hilbert space and should answer the question of how much $\hat{\mathcal{O}}(t)$ deviates from the original one $\hat{\mathcal{O}}(0)$ as it evolves in time. Specifically, as $\hat{\mathcal{O}}(t)$ evolves in time, it will be described by a linear combination of $\mathcal{L}^n \hat{\mathcal{O}}$, and we need to construct an orthonormalized basis to measure the spread of $\hat{\mathcal{O}}(t)$. To do so, we introduce an inner product $(\hat{A}\vert\hat{B})$ between operators $\hat{A}$ and $\hat{B}$, and then do orthogonalization by the Gram-Schmidt method. By choosing a suitable inner product, a matrix $L_{m,n}$, defined by
\begin{align}
L_{m,n}\equiv (\hat{O}_m|\mathcal{L}\hat{O}_n),
\end{align}
becomes a Hermitian matrix, where $\{ \hat{\mathcal{O}}_n \}$ is the obtained orthonormalized basis such that $(\hat{\mathcal{O}}_m\vert\hat{\mathcal{O}}_n)=\delta_{mn}$.
We call $\{ \hat{\mathcal{O}}_n \}$ Krylov basis, which obeys 
\begin{align}
&\hat{\mathcal{O}}_{-1}\equiv 0,\ \ \hat{\mathcal{O}}_0\equiv \hat{\mathcal{O}},\\
&\mathcal{L}\hat{\mathcal{O}}_n=a_n\hat{\mathcal{O}}_n+b_n\hat{\mathcal{O}}_{n-1}+b_{n+1}\hat{\mathcal{O}}_{n+1} ,\label{anbnOn}\\
&L_{m,n}\equiv (\hat{O}_m|\mathcal{L}\hat{O}_n)=
\left(
\begin{array}{ccccc}
a_0 & b_1 & 0 & 0 & \cdots \\
b_1 & a_1 & b_2 & 0 & \cdots \\
0 & b_2 & a_2 & b_3 & \cdots \\
0 & 0 & b_3 & a_3 & \cdots \\
\vdots & \vdots & \vdots & \vdots & \ddots
\end{array}
\right),
\end{align}
where $a_n$ and $b_n$ are called Lanczos coefficients and play a very important role in the calculation of the Krylov complexity.\\

In particular, these Lanczos coefficients can be read off from a two-point correlation function. The method is as follows. First, consider a two-point correlator $G(t)=(\hat{\mathcal{O}}(t)|\hat{\mathcal{O}})$, where its Taylor expansion is given by
\begin{align}
G(t)=\sum_{n=0}M_n \frac{(-it)^n}{n!},\ M_n=\left. \frac{1}{(-i)^n}\frac{d^n G(t)}{dt^n}\right|_{t=0}=(\hat{\mathcal{O}}_0|\mathcal{L}^n\hat{\mathcal{O}}_0)\equiv(\hat{\mathcal{O}}_0|\mathcal{L}^n|\hat{\mathcal{O}}_0).
\end{align}
Here, $M_n$ can also be computed from the Fourier transformation of $G(t)$ as
\begin{align}
&M_n=\int_{-\infty}^{\infty} \frac{d\omega}{2\pi}\omega^n f(\omega),\\
&f(\omega) \equiv \int_{-\infty}^{\infty}dt e^{i\omega t}G(t).
\end{align}
Since $\mathcal{L}^n\hat{\mathcal{O}}_0$ can be expressed by $\{ \hat{\mathcal{O}}_n \}$ and the Lanczos coefficients from (\ref{anbnOn}), 
the Lanczos coefficients can be read from $M_n$ as follows.
\begin{align}
M_1&=(\hat{\mathcal{O}}_0|\mathcal{L}|\hat{\mathcal{O}}_0)=a_0,\\
M_2&=(\hat{\mathcal{O}}_0|\mathcal{L}^2|\hat{\mathcal{O}}_0)=a_0^2+b_1^2,\\
M_3&=(\hat{\mathcal{O}}_0|\mathcal{L}^3|\hat{\mathcal{O}}_0)=a_0^3+2a_0b_1^2+a_1b_1^2,\ \  \cdots .
\end{align}

Once the Lanczos coefficients are obtained, the Krylov complexity can then be calculated as follows. First, we prepare 
\begin{align}
\hat{\mathcal{O}}(t)\equiv \sum_{n=0} i^n \varphi_n(t) \hat{\mathcal{O}}_n.
\end{align}
This is simply a rewriting of the time evolution of $\hat{\mathcal{O}}$, where $i^n \varphi(t)$ is the coefficient when we expand $\hat{\mathcal{O}}(t)$ in the Krylov basis. In other words,
\begin{align}
\varphi_n(t)=i^{-n} (\hat{\mathcal{O}}_n|\hat{\mathcal{O}}(t)) .
\end{align}
On the other hand, time derivative of $\hat{\mathcal{O}}(t)$ is
\begin{align}
\frac{d\hat{\mathcal{O}}(t)}{dt}&= \sum_{n=0} i^n \frac{d\varphi_n(t)}{dt} \hat{\mathcal{O}}_n\notag \\
&=i[H, \hat{\mathcal{O}}(t)]=i\mathcal{L}\hat{\mathcal{O}}(t)=\sum_{n=0} i^{n+1} \varphi_n(t) \mathcal{L} \hat{\mathcal{O}}_n,
\end{align}
and
\begin{align}
\frac{d\varphi_n(t)}{dt}=ia_n\varphi_n(t)-b_{n+1}\varphi_{n+1}(t)+b_{n}\varphi_{n-1}(t),\\
\varphi_{-1}\equiv 0,\ \ \varphi_0(t)=G^*(t).
\end{align}
The Krylov complexity is defined by
\begin{align}
K(t)\equiv \sum_{n=1}^{\infty} n |\varphi_n(t)|^2.
\end{align}
As is clear from the definition and its construction, the Krylov complexity represents the operator growth in the Krylov subspace of total Hilbert space associated with the time evolution of an initial local operator.
This quantity is often used to study chaotic systems.

From the above, it can be seen that once the Lanczos coefficients are known, the Krylov complexity can be specifically calculated from them.  Furthermore, its behavior at the late time is found to be determined from the asymptotic form of the Lanczos coefficients \cite{Barbon:2019wsy}. In particular, when the system is chaotic, $b_n$ is expected to increase linearly with $n$ when $n$ is large, which leads to an exponential growth of the Krylov complexity
\begin{align}
b_n \sim \alpha n,\ \ K(t)\sim e^{2\alpha t},
\end{align}
up to log corrections.
Furthermore, the asymptotic form of $b_n$ is shown in \cite{Lubinsky1988} to reflect the tail structure of the spectrum density $f(\omega)$. For example, if $f(\omega) \sim \exp [-|\omega / \omega_0|]$ at large $\vert\omega\vert$, the asymptotic form of $b_n$ is $b_n \sim \frac{\pi \omega_0}{2}n$ and behaves linearly for sufficiently large $n$. This observation leads to condition B in this paper.

\section{Heat Kernel}\label{App1}
In this appendix, we give a rough calculation of the heat kernel.\\

\noindent{\bf Heat Kernel on $\mathbb{S}^1$}\\
From the definition,
\begin{align}
K_{\mathbb{S}^1}(s,\tau) & = \braket{\tau| e^{s\partial_{\tau}^2}|0} = \sum_{n,m} \braket{\tau|n}\braket{n|e^{s\partial_{\tau}^2}|m}\braket{m|0}
\end{align}
Now wave function is $\braket{\tau|n} = \frac{1}{\sqrt{\beta}} e^{ - i p_n \tau},\ \ (p_n = \frac{2\pi n}{\beta})$. Therefore
\begin{align}
K_{\mathbb{S}^1}(s,\tau) & = \frac{1}{\beta} \sum_{n,m}  e^{- i p_n \tau} e^{-sp_n^2}\delta_{nm} = \frac{1}{\beta} \sum_{n} e^{-sp_n^2 - i p_n \tau}\notag\\
& = \frac{1}{\sqrt{4\pi s}}\sum_n e^{-\frac{(\tau +n\beta)^2}{4s}}
\end{align}
In the last equality, Poisson resummation 
\begin{align}
\sum_{n \in \mathbb{Z}} f(n) = \sum_{k \in \mathbb{Z}} f(k),\ \ f(k) = \int_{-\infty}^{\infty} f(x) e^{-i2\pi k x}dx
\end{align}
was used.\\

\noindent{\bf Heat Kernel on $\mathbb{S}^3$}\\
Eigenvalue $\lambda_{\ell}$ of $\nabla_{\mathbb{S}^3}$ where $\mathbb{S}^3$ has the radius $R$ and it's degeneracy $D_{\ell}$ is known,
\begin{align}
\lambda_{\ell} = -\frac{\ell(\ell+2)}{R^2},\ \ D_{\ell} = (\ell+1)^2,\ \ \ell=0,1,2,\cdots
\end{align}
Up to constant factor,
\begin{align}
K_{\mathbb{S}^3}(s) &\propto \sum_{\ell=0}^{\infty} (\ell+1)^2e^{-s\ell(\ell+2)/R^2} = \sum_{n=1}^{\infty} n^2e^{-s(n^2-1)/R^2} \propto \sum_{n=-\infty}^{\infty} n^2e^{-s(n^2-1)/R^2}\notag \\
&\propto \frac{e^{s/R^2}}{s^{3/2}}\sum_{\ell \in \mathbb{Z}} e^{-\frac{\pi^2 R^2\ell^2 }{s}}\left( 1- 2 \frac{\pi^2 R^2 \ell^2}{s}\right)
\end{align}
Finally, Poisson resummation was again used. To take even the coefficients into account, we have
\begin{align}
K_{\mathbb{S}^3}(s) & =\sum_{l,k,m} |\braket{\theta_1,\theta_2,\theta_3|\ell,k,m}|^2 e^{-s\ell(\ell+2)/R^2} 
\end{align}
There is a good formula, Uns\"{o}ld's theorem. If we consider  Super spherical harmonics on $\mathbb{S}^{n}$ which has $n$ angular coordinate $\omega_1,\omega_2,\cdots, \omega_n$, then those have $n$ quantum number $l_1,l_2,\cdots,l_n$,
\begin{align}
\sum_{k,m} |\braket{\omega_1,\omega_2,\cdots, \omega_n|l_1,l_2,\cdots,l_n}|^2 = \frac{D_{\ell}}{{\rm Vol}(\mathbb{S}^{n})}
\end{align}
If we consider $n=2$ case, $|\braket{\theta,\phi,|l,m}|^2=|Y_{\ell m }(\theta,\phi)|^2$, $D_{\ell} = 2\ell+1$ and ${\rm Vol}(\mathbb{S}^2) = 4\pi$, therefore
\begin{align}
\sum_{m=-\ell}^{\ell} |Y_{\ell m }(\theta,\phi)|^2 = \frac{2\ell+1}{4\pi},\ \ {\rm on\ Unit\ Sphere}
\end{align}
This is a well-known result. Now we want use $n=3$ case, then $|\braket{\omega,\theta,\phi,|l,k,m}|^2=|Y_{\ell k m }(\omega,\theta,\phi)|^2$, $D_{\ell} = (\ell+1)^2$ and ${\rm Vol}(\mathbb{S}^2) = 2\pi^2 R^3$. Therefore, we get
\begin{align}
\sum_{k,m} |\braket{\omega,\theta,\phi,|l,k,m}|^2 = \frac{(\ell+1)^2}{2\pi^2 R^3}
\end{align}
Then
\begin{align}
K_{\mathbb{S}^3}(s) & = \sum_{l=0}^{\infty} \frac{(\ell+1)^2}{2\pi^2 R^3} e^{-s\ell(\ell+2)/R^2} = \frac{1}{2}\sum_{l\in \mathbb{Z}} \frac{n^2}{2\pi^2 R^3} e^{-s(n^2-1)/R^2} \notag\\
&= \frac{e^{s/R^2}}{(4\pi s)^{3/2}}\sum_{\ell \in \mathbb{Z}} e^{-\frac{\pi^2 R^2\ell^2 }{s}}\left( 1- 2 \frac{\pi^2 R^2 \ell^2}{s}\right).
\end{align}

\bibliography{paper}
\bibliographystyle{utphys}

\end{document}